\pdfoutput=1
\documentclass[10pt,conference,letterpaper]{IEEEtran}

\usepackage[utf8]{inputenc}
\usepackage[T1]{fontenc}

\usepackage{algorithmic}
\usepackage{algorithm}
\usepackage{url}
\usepackage{color}
\usepackage{listings}
\usepackage{microtype}

\usepackage{tikz}
\usepackage{tkz-euclide}
\usetikzlibrary{arrows,positioning,shapes.geometric}
\usepackage{pgfplots}

\usepackage{amsmath}
\usepackage{amssymb}
\usepackage{soul}
\usepackage{calc}
\usepackage{multirow}
\usepackage{booktabs}
\usepackage{xspace}
\usepackage[binary-units]{siunitx}
\usepackage[caption=false]{subfig}
\usepackage{rotating}
\usepackage{enumitem}
\usepackage{setspace}
\usepackage{hyperref}
\usepackage{placeins}

\usepackage{xcolor, colortbl}
\definecolor{Gray}{gray}{0.85}
\newcolumntype{a}{>{\columncolor{Gray}}c}

\setlist[enumerate]{topsep=1pt,itemsep=-1ex,partopsep=1ex,parsep=1ex}

\newcommand{\mpisend}{\texttt{MPI\_Send}\xspace}

\newcommand{\mpiirecv}{\texttt{MPI\_Irecv}\xspace}

\newcommand\mpibcast{\texttt{MPI\_Bcast}\xspace}
\newcommand\mpiscatter{\texttt{MPI\_Scatter}\xspace}
\newcommand\mpiscatterv{\texttt{MPI\_Scatterv}\xspace}
\newcommand\mpigather{\texttt{MPI\_Gather}\xspace}

\newcommand\mpiallgather{\texttt{MPI\_Allgather}\xspace}

\newcommand\mpialltoall{\texttt{MPI\_Alltoall}\xspace}

\newcommand\mpireduce{\texttt{MPI\_Reduce}\xspace}
\newcommand\mpiallreduce{\texttt{MPI\_Allreduce}\xspace}
\newcommand\mpireducescatter{\texttt{MPI\_Reduce\_scatter}\xspace}
\newcommand\mpireducescatterblock{\texttt{MPI\_Reduce\_scatter\_block}\xspace}
\newcommand\mpiscan{\texttt{MPI\_Scan}\xspace}
\newcommand\mpiexscan{\texttt{MPI\_Exscan}\xspace}
\newcommand\mpireducelocal{\texttt{MPI\_Reduce\_local}\xspace}

\newcommand{\productname}[1]{#1} 
\newcommand{\code}[1]{\texttt{#1}}

\newcommand{\skampi}{\productname{SKaMPI}\xspace}
\newcommand{\reprompi}{\productname{ReproMPI}\xspace}

\newcommand{\mpicroscope}{\productname{mpicroscope}\xspace}

\newcommand{\openmpi}{\productname{Open\,MPI}\xspace}

\newcommand{\mvapich}{\productname{MVAPICH}\xspace}
\newcommand{\infiniband}{IB\xspace}
\newcommand{\gcc}{gcc\xspace}
\newcommand{\mpirun}{\texttt{mpirun}\xspace}
\newcommand{\mpiruns}{\texttt{mpirun}s\xspace}

\newcommand{\jupitermvapich}{\productname{MVAPICH2-2.1}\xspace}
\newcommand{\jupiternecmpi}{\productname{NEC\,MPI~1.3.1}\xspace}
\newcommand{\jupiteropenmpi}{\productname{Open\,MPI~1.10.1}\xspace}

\newcommand{\vscintelmpi}{\productname{Intel\,MPI~5.0~(Update~3)}\xspace}

\newcommand{\fermibgqmpi}{\productname{Blue\,Gene/Q\,MPI}\xspace}

\newcommand{\mpi}{MPI\xspace}

\newcommand{\fig}{Figure\xspace}
\newcommand{\Fig}{Fig.\xspace}

\newcommand{\tab}{Table\xspace}

\newcommand{\append}{Appendix\xspace}

\newcommand{\Sect}{Section\xspace}

\newcommand{\eg}{e.g.\xspace}
\newcommand{\ie}{i.e.\xspace}
\newcommand{\etal}{et~al.\xspace}

\newcommand{\cf}{cf.\xspace}

\newcommand{\runtime}{run-time\xspace}
\newcommand{\runtimes}{run-times\xspace}

\newcommand{\pgmpi}{PGMPI\xspace}

\newcommand{\Bytes}{Bytes\xspace}
\newcommand{\Byte}{Byte\xspace}

\newcommand{\machone}{\emph{Jupiter}\xspace}
\newcommand{\machtwo}{\mbox{\emph{VSC-3}}\xspace}
\newcommand{\machthree}{\mbox{\emph{Fermi}}\xspace}

\newcommand{\testwilcox}{\emph{Wilcoxon rank-sum test}\xspace}
\newcommand{\testkolm}{\emph{Kolmogorov-Smirnov test}\xspace}

\newcommand{\nmpiruns}{\ensuremath{R}\xspace}
\newcommand{\nrep}{\ensuremath{r_i}\xspace}

\newcommand{\funci}{\ensuremath{f}\xspace}
\newcommand{\msize}{\ensuremath{m_i}\xspace}
\newcommand{\np}{\textit{p}\xspace}

\newcommand{\repRSE}{\ensuremath{RSE}\xspace}
\newcommand{\repCOVMEAN}{\ensuremath{COV_{mean}}\xspace}
\newcommand{\repCOVMEDIAN}{\ensuremath{COV_{median}}\xspace}

\newcommand{\threshRSE}{\ensuremath{t_{\repRSE}}}
\newcommand{\threshCOVMEAN}{\ensuremath{t_{\repCOVMEAN}}}
\newcommand{\threshCOVMEDIAN}{\ensuremath{t_{\repCOVMEDIAN}}}

\newcommand{\winCOVMEAN}{\ensuremath{w_{\repCOVMEAN}}}
\newcommand{\winCOVMEDIAN}{\ensuremath{w_{\repCOVMEDIAN}}}

\newcommand{\stepone}{\emph{Step~NREP}\xspace}
\newcommand{\steptwo}{\emph{Step~MEASURE}\xspace}
\newcommand{\stepthree}{\emph{Step~ANALYZE}\xspace}

\newcommand{\guidelt}{\ensuremath{\preceq}\xspace}

\newcommand{\glmono}{monotony\xspace}
\newcommand{\glsplit}{split-robustness\xspace}
\newcommand{\glpattern}{pattern\xspace}

\newenvironment{tightcenter}{%
  \setlength\topsep{0pt}
  \setlength\parskip{0pt}
  \begin{center}
}{%
  \end{center}
}

\colorlet{punct}{red!60!black}
\definecolor{background}{HTML}{EEEEEE}
\definecolor{delim}{RGB}{20,105,176}
\colorlet{numb}{magenta!60!black}

\lstdefinelanguage{json}{
    stepnumber=1,
    showstringspaces=false,
    breaklines=true,
    frame=lines,
    literate=
     *{0}{{{\color{numb}0}}}{1}
      {1}{{{\color{numb}1}}}{1}
      {2}{{{\color{numb}2}}}{1}
      {3}{{{\color{numb}3}}}{1}
      {4}{{{\color{numb}4}}}{1}
      {5}{{{\color{numb}5}}}{1}
      {6}{{{\color{numb}6}}}{1}
      {7}{{{\color{numb}7}}}{1}
      {8}{{{\color{numb}8}}}{1}
      {9}{{{\color{numb}9}}}{1}
      {:}{{{\color{punct}{:}}}}{1}
      {,}{{{\color{punct}{,}}}}{1}
      {\{}{{{\color{delim}{\{}}}}{1}
      {\}}{{{\color{delim}{\}}}}}{1}
      {[}{{{\color{delim}{[}}}}{1}
      {]}{{{\color{delim}{]}}}}{1},
}

\makeatletter
\def\lst@makecaption{%
  \def\@captype{table}%
  \@makecaption
}
\makeatother

\floatname{algorithm}{Listing}
\newtheorem{definition}{Definition}

\IEEEoverridecommandlockouts

\begin{document}

\title{PGMPI: Verifying Self-Consistent MPI~Performance Guidelines}

\author{\IEEEauthorblockN{Sascha Hunold, Alexandra Carpen-Amarie,
    Felix Donatus L\"ubbe, and Jesper
    Larsson Tr\"aff} 
  
    \IEEEauthorblockA{
    TU Wien\\
    Faculty of Informatics, Institute of Information Systems\\
    Research Group for Parallel Computing,\\
    Favoritenstrasse 16/184-5,  1040 Vienna, Austria\protect\\
    Email: \{hunold,carpenamarie,luebbe,traff\}@par.tuwien.ac.at}
  \thanks{This work was supported by the  Austrian Science Fund (FWF): P26124 and P25530}
}

\maketitle

\renewcommand\thelstlisting{\arabic{lstlisting}}

\begin{abstract}
  The Message Passing Interface (\mpi) is the most commonly used
  application programming interface for process communication on
  current large-scale parallel systems. Due to the scale and
  complexity of modern parallel architectures, it is becoming
  increasingly difficult to optimize \mpi libraries, as many factors
  can influence the communication performance. To assist \mpi
  developers and users, we propose an automatic way to check whether
  \mpi libraries respect self-consistent performance guidelines for
  collective communication operations. We introduce the \pgmpi
  framework to detect violations of performance guidelines through
  benchmarking.  Our experimental results show that \pgmpi can
  pinpoint undesired and often unexpected performance degradations of
  collective \mpi operations. We demonstrate how to overcome
  performance issues of several libraries by adapting the algorithmic
  implementations of their respective collective \mpi calls.
\end{abstract}


\section{Introduction}
\label{sec:intro}

Communication libraries implementing the Message Passing Interface
(\mpi) are major building blocks for developing parallel, distributed,
and large-scale applications for current supercomputers. The
performance of parallel codes is therefore highly dependent on the
efficiency of \mpi implementations. Much research is currently
conducted to cope with the problems of exascale computing in MPI.

Assessing the performance of MPI implementations is vital for
developers, vendors, and users of the libraries. However, the
performance of MPI libraries can be measured in different ways.  A
common approach is to run a set of \mpi micro-benchmarks, such as
\skampi~\cite{ReussnerST02} or
\reprompi\footnote{\url{https://github.com/hunsa/reprompi}}~\cite{tpds2016}. Micro-benchmarks
usually report the measured (mean or median) \runtime of a given MPI
function for different message sizes, \eg, the \runtime of \mpibcast
for broadcasting a \SI{1}{\Byte} message.  Developers can gain
insights on how the \runtime of an MPI function depends on the message
size for a fixed number of processes.  It is also possible to assess
the scalability of MPI functions when the number of processes is
increased and the message size stays fixed.

Verifying self-consistent \mpi performance guidelines is an
alternative, orthogonal method for analyzing the performance of \mpi
libraries~\cite{TraffGT10}. This approach does not require explicit
performance models. Instead, performance guidelines form a set of
rules that an MPI library is expected to fulfill. A performance
guideline usually defines an upper bound on the \runtime behavior of a
specialized MPI function. For example, one performance guideline
states that a call to \mpiscatter of $n$ data elements should ``not be
slower'' than a call to \mpibcast with $n$ data elements, as the
semantics of an \mpiscatter operation could be emulated using
\mpibcast~\cite{TraffGT10}. Only minor efforts have been made to
systematically test self-consistent performance guidelines for MPI
implementations in practice, one example being the \mpicroscope
benchmark~\cite{Traff12}.  To close this gap, we introduce the
benchmarking framework \pgmpi that can automatically verify
performance guidelines of \mpi libraries.

We make the following contributions: (1) We propose the benchmarking
framework \pgmpi to detect performance-guideline violations. (2) We
present a systematic, experimental verification of performance
guidelines for several \mpi libraries. (3) We examine different use
cases, for which the detection of guideline violations enabled us to
tune and improve the libraries' performance.

In \Sect~\ref{sec:notation}, we state the scientific problem and
introduce our notation. We continue with summarizing related work and
comparing it to our approach in \Sect~\ref{sec:rel_work}.  We
introduce the \pgmpi framework in \Sect~\ref{sec:pgmi_framework} and
present an experimental evaluation of different MPI libraries using it
in \Sect~\ref{sec:exp_results}. We summarize our findings and conclude
in \Sect~\ref{sec:conclusions}.

\section{Problem Statement and Notation}
\label{sec:notation}

Träff~\etal~\cite{TraffGT10} introduced self-consistent performance guidelines for MPI
libraries as follows: The \runtime of two MPI functionalities $A$ and
$B$ can be ordered using the relation \guidelt as
$\texttt{MPI}\_A(n) \guidelt \texttt{MPI}\_B(n)$, which means that
functionality $\texttt{MPI}\_A(n)$ is possibly faster than
functionality $\texttt{MPI}\_B(n)$ for (almost) all communication
amounts $n$.  Performance guidelines are defined for
a fixed number of processes~\np, and thus, they do not mention \np
explicitly.  However, the communication volume per process may vary
depending on the semantics of a given MPI function and the number of
processes. For example, in the case of \mpibcast, the total size~$n$
is equal to the message size being transferred to each process. In
contrast, the individual message size for an \mpiscatter is a fraction
of the total communication volume $n$, \ie, each process receives
$n/p$ elements.

We examine three types of performance guidelines: (1)~\emph{\glmono},
(2)~\emph{\glsplit}, and (3)~\emph{\glpattern}.  The monotony
guideline
\begin{tightcenter}
$\texttt{MPI}\_A(n) \guidelt \texttt{MPI}\_A(n + k)$  
\end{tightcenter}
ensures that communicating a larger volume should not decrease the
communication time.  

\noindent
The split-robustness guideline
\begin{tightcenter}
\texttt{MPI}\_A(n) \guidelt k\, \texttt{MPI}\_A(n/k)
\end{tightcenter}
states that communicating a total volume of $n$ data elements should
not be slower than sending $\frac{n}{k}$ elements in $k$ steps.

Pattern guidelines define upper bounds on the performance of MPI
communication operations. The idea is that a specialized MPI function
should not have a larger running time than a combination of other MPI
operations, which emulate the functionality of the specialized
function. Let us consider the following \emph{pattern} performance
guidelines:
\begin{align*}
\mpiscatter(n) & \guidelt \mpibcast(n) \, , \, \mbox{and} \\
\mpibcast(n) & \guidelt \mpiscatter(n) \\
             &  + \mpiallgather(n) \quad . 
\end{align*}
The first states that \mpiscatter should not be slower than \mpibcast.
The reason is that the semantics of \mpiscatter can be implemented
using \mpibcast, by broadcasting the entire vector before processes
take their share depending on their rank.  The second guideline states
that a call to \mpibcast should be at least as fast as a combination
of \mpiscatter and \mpiallgather, which we call the \emph{mock-up}
version of \mpibcast, as it emulates its semantics~\cite{ChanHPG07}.

\begin{figure}[t]
  \centering
  \begin{tikzpicture}[scale=0.50]
    \Large
    \begin{axis}[
      height=0.5\textwidth,
      width=.9\textwidth,
      grid=both,
      xlabel={message size [Bytes]},
      ylabel={\runtime [$\mu$s]},
      xtick={2,4,8,16,32,64},
      ylabel style={yshift=0.0cm},
      xlabel style={yshift=.1cm},
      legend style={at={(axis cs:1,50)},anchor=south west}
      ]
      \addplot[mark=*,blue] coordinates {(2,10) (4,14) (8,18) (10, 24) (12, 23) (16,25) (32,60)};
      \addplot[mark=*,color={rgb:red,4;green,2;yellow,1}] coordinates {(2,12) (4,16) (8,22) (10, 30) (12, 35) (16,40) (32,54)};
      \addplot[mark=+, red, thick] coordinates {(10,24) (12, 23)};
      \node[anchor=south, red] at (axis cs:16,14) {(1) monotony violation};

      \addplot[mark=+, red, thick, dashed] coordinates {(16,25) (32,50)};
      \addplot[mark=+, red, thick] coordinates {(16,25) (32,60)};
      \node[anchor=south, red] at (axis cs:26,23) {(2) split-robustness violation};

      \addplot[mark=+, red, thick] coordinates {(32,60) (32,54)};
      \node[anchor=south, red] at (axis cs:23,54) {(3) pattern violation};

      \addlegendentry{\mpigather}
      \addlegendentry{\mpiallgather}
    \end{axis}
  \end{tikzpicture}
  \caption{\label{fig:example_violations}Possible violations of performance guidelines.}
\end{figure}
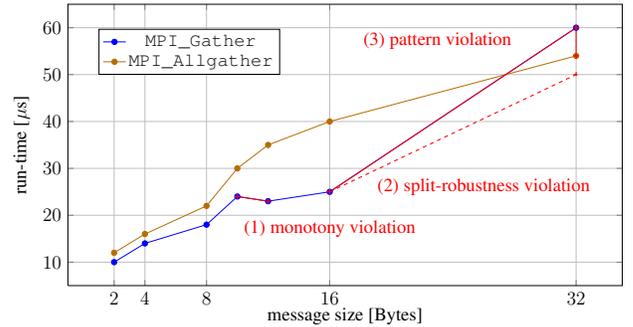

\fig~\ref{fig:example_violations} depicts the described
violations. The \emph{monotony} is violated when a communication
operation becomes faster when the communication volume increases. A
\emph{split-robustness} violation occurs when sending a larger message
in multiple chunks (several smaller messages) is faster than sending
only one large message.  Last, a \emph{pattern} violation denotes the
case when a specific, specialized MPI communication can be emulated by
other MPI communication operations, and when this emulation is faster
than the specialized version. For example, \mpigather can be emulated
using \mpiallgather, and thus, the \runtime of \mpiallgather should
not be faster than \mpigather.

\section{Related Work}
\label{sec:rel_work}

Collective communication operations are a central part of the MPI
standard, as they are essential for many large-scale
applications. Chan~\etal~\cite{ChanHPG07} provide an overview of
typical, blocking collectives and their implementations, as well as
lower bounds for the communication cost of each function. For
different network topologies, the authors devise algorithms that
achieve the lower bounds for either the latency or the bandwidth
component. As the model of parallel computation in this paper is
rather simplistic, we aim to complement this study by carefully
benchmarking MPI collectives on actual hardware.

Träff~\cite{Traff12} proposed the MPI benchmark \mpicroscope, which
can verify two self-consistent performance guidelines:
``split-robust'' and ``monotone''. In the present work, we extend this
functionality by testing various pattern violations, using the
experimental framework that was proposed by
Hunold~\etal~\cite{tpds2016} for better reproducibility of the
experimental results.  While we focus on performance guidelines for
collectives, previous works have also formulated performance
guidelines for derived datatypes~\cite{Gropp:2011} and MPI-IO
operations~\cite{GroppKRTT08}.

As hardware and software factors can influence the performance of MPI
collectives, tuning MPI parameters is an essential part for achieving
high performance, when installing an MPI library. Yet, optimizing and
tuning MPI operations are orthogonal steps compared to the
verification of self-consistent performance guidelines, \ie, the
latter can help us to verify whether \runtimes of collectives are
consistent in terms of expected performance.  For example, the guidelines can be used to
ensure that Gather is faster than Allgather for the same problem
size. For that reason, if a violation occurs, it usually means that one
collective can be tuned. To optimize the latency of collectives at
\runtime, one can employ the STAR-MPI routines~\cite{FarajYL06}.
When a call to a specific MPI function is issued, STAR-MPI selects one
of the available algorithms and measures its \runtime. When STAR-MPI
has enough knowledge about the performance of different algorithms, it
is able to pick a good algorithm for a specific case.

Selecting the right algorithm to implement a given MPI function is
only one step towards tuning MPI libraries. Another problem is finding
the right parameter settings that \runtime systems of MPI libraries
like \openmpi or \mvapich offer. Chaarawi~\etal~\cite{ChaarawiSGF08}
introduced the OTPO tool that can be used to tune \openmpi \runtime
parameters. OTPO takes as input the \runtime parameters to be tuned as
well as their respective ranges, and then starts measuring for all
combinations of parameter values.  Another approach to tune \openmpi
parameters has been proposed by
Pellegrini~\etal~\cite{PellegriniWFM09}, where the parameter values
are predicted using machine learning techniques.

The performance guidelines are formulated as a function of the
communication volume. It is also possible to examine the scalability
of MPI collectives when increasing the number of processes.
Shudler~\etal~\cite{Shudler15} proposed a framework to compare
performance characteristics of HPC applications with a theoretical
performance model.
The framework fits the recorded benchmarking data to analytic speedup
functions and compares the experimentally determined scalability
behavior to this expected performance model. A model mismatch
indicates a scalability problem of the parallel code section.

\section{\pgmpi: Verifying MPI Performance Guidelines}
\label{sec:pgmi_framework}

We now introduce the \pgmpi framework to verify self-consistent
performance guidelines of MPI libraries. In the first step, \pgmpi
experimentally determines the number of repetitions needed to obtain
stable, reproducible \runtime measurements (\stepone). In the second
step, the framework performs \runtime measurements of all functions
for which performance guidelines are formulated (\steptwo). The data
analysis and the statistical verification of performance guidelines is
carried out in the last step (\stepthree).

\subsection{Obtaining Reproducible Results}

We start by looking at the main (second) step of \pgmpi (\steptwo), in
which the \runtimes of MPI functions and their emulating counterparts
are measured. The guideline-checking program takes as input a set of
pattern guidelines, 
each defined by a pair consisting of an MPI function and its emulating
mock-up function.
Our \pgmpi framework will measure the \runtime of one of the specified
MPI functions~\funci for all given message sizes~\msize and the number
of processes~\np that are given in the input file. Within one call to
\mpirun, each individual measurement for (\funci, \msize) is repeated
\nrep~times, where $r_i$ is defined for each \msize.  As we expect
that mean (or median) \runtimes vary between different calls to
\mpirun~\cite{tpds2016}, the \pgmpi framework measures the \runtime of
each MPI function~\funci over \nmpiruns \mpiruns.

\subsection{Determining the Number of Repetitions}

A major problem in \mpi benchmarking is the question of how long (how
many times) to measure. We need to find the right trade-off between
time and measurement stability.
One way of dealing with this problem is by executing the experiment
sufficiently often, \eg, \num{1000} times. This would alleviate the
problem of low measurement stability, but most often, we cannot afford
long-running benchmarking experiments. Therefore, we formulate the
following problem:

\begin{definition}
  The NREP problem is to find a suitable number of repetitions~\nrep
  for the tuple $(\funci, \msize, \np)$, such that the obtained
  \runtime metric after \nrep repetitions of function \funci with
  \msize\Bytes on \np processes is reproducible between different
  calls to \mpirun. Reproducible in this case means that the
  distribution of the measured values (for a specific metric) obtained
  from \nmpiruns \mpiruns has a small variance.
\end{definition}

\begin{figure*}[t]
  \centering
  \includegraphics[width=.9\linewidth]{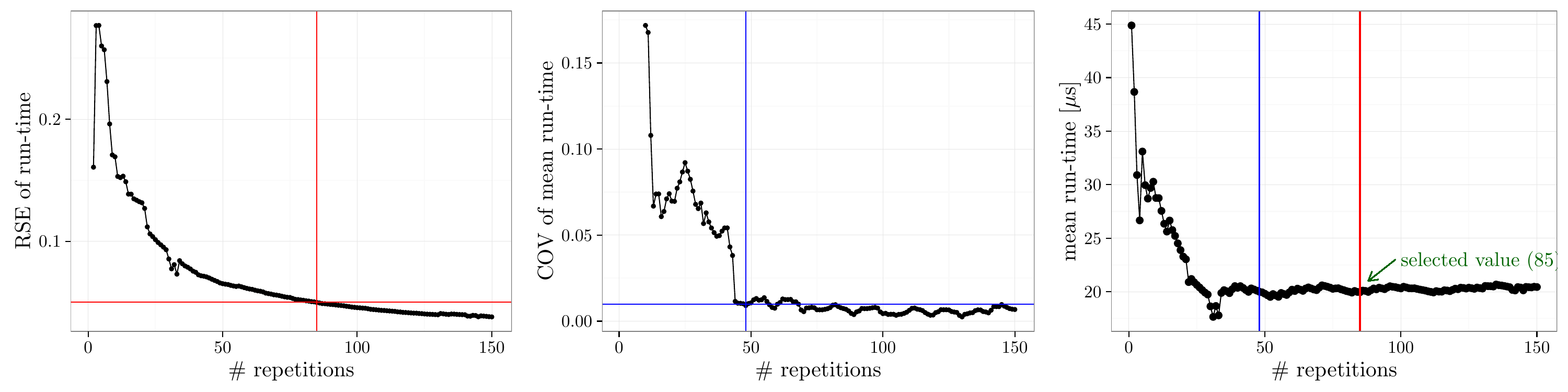}  
  \caption{Example of estimating the required number of repetitions
    for \mpiallgather (\SI{16}{\byte}, \num{16x1} processes, \machone,
    $\threshRSE=\num{0.025}$, $\threshCOVMEAN=\num{0.01}$,
    $\winCOVMEAN=\num{20}$).}
  \label{fig:nrep_estimation}
\end{figure*}

We have experimented with various ways of estimating the number of
repetitions needed to obtain reproducible results. One possibility is
to monitor the relative standard error of the mean (\repRSE). \skampi,
for example, stops the measurements when the \repRSE falls below a
threshold of \num{0.1}~\cite{ReussnerST02}. Although we have tested
many different ways to solve the NREP problem, we could not find a
generally superior approach. We therefore designed the NREP predictor
for \stepone of the \pgmpi framework in a flexible manner.  The
framework currently provides three different methods (metrics) for
solving the NREP problem, but new metrics can be added. The NREP
prediction may stop
\begin{enumerate}
\item when the \textbf{relative standard error} (\repRSE) is smaller
  than some predefined threshold \threshRSE; or
\item when the \textbf{coefficient of variation of the mean \runtime}
  (\repCOVMEAN) is smaller than some predefined threshold
  \threshCOVMEAN. The value of the \repCOVMEAN is computed over the
  last \winCOVMEAN means (window size); or
\item when the \textbf{coefficient of variation of the median
    \runtime} (\repCOVMEDIAN) is smaller than some predefined
  threshold \threshCOVMEDIAN using a window size of \winCOVMEDIAN.
\end{enumerate}
Users can choose the NREP prediction method on the command line as follows:
\begin{lstlisting}[language=bash,basicstyle=\small\ttfamily,frame=lines]
mpirun -np 4  ./mpibenchmarkPredNreps \
  --calls-list=MPI_Reduce --msizes-list=8 \ 
  --rep-prediction min=20,max=1000,step=10 \
  --pred-method=rse --var-thres=0.025
\end{lstlisting}
It is also possible to combine different metrics, \ie, the NREP
prediction stops when all selected metrics have been positively
evaluated. An example is shown in \Fig~\ref{fig:nrep_estimation}, in
which both the \repRSE and the \repCOVMEAN need to be below a specific
threshold (marked with horizontal lines). The prediction function for
the \repRSE metric stops after \num{85} iterations, at which the
\repCOVMEAN value is also below its threshold. As a result, \num{85}
is the number of iterations that will be used when collecting
benchmark data in \steptwo. To cope with the \runtime variation
between different \mpiruns, we perform three NREP predictions for each
message size and select the maximum number of repetitions obtained.

\subsection{Statistically Verifying Performance Guidelines}
\label{sec:verifying_guidelines}

After gathering the measurement results, \pgmpi can proceed to
\stepthree, which consists of the data processing and the verification
of performance guidelines. We now explain which statistical methods
are applied for guideline verification. For each MPI function, for
which guidelines were formulated, the experimental results comprise
\nmpiruns (number of \mpiruns) data sets for a specific number of
processes~$\np$. Each data set contains \nrep \runtime measurements
for a specific message size \msize. We first reduce the number of
measurements per tuple ($\mpirun_j$, \msize, $\np$) to a single value,
by computing the median \runtime over the \nrep measurements. In this
way, we obtain a distribution of \nmpiruns medians (median \runtimes)
for each message size \msize and processes~\np.  The various
performance guidelines will then be verified using these distributions
of medians.

\subsubsection{Monotony Guideline}

\pgmpi checks for each pair of adjacent message sizes $m_i$ and $m_j$,
$m_i<m_j$ that the \runtime of an MPI function with a message size of
$m_i$ is not larger than the \runtime with a size of $m_j$.
We use the \testwilcox~\cite{hollander2014} to test whether the distribution of medians at
$m_i$ is smaller or equal than the one at $m_j$. If the test rejects
our hypothesis, we have statistical confidence (at the provided
confidence level) that the monotony between message sizes $m_i$ and
$m_j$ is violated.

\subsubsection{Split-Robustness Guideline}

We want to verify that sending a message of size~$m_j$ by transferring
$k$ packets of size $m_i < m_j$ is not faster than sending only one
message of size $m_j$. We are only given the \runtime distribution of
one MPI function at $m_i$. Unfortunately, we have no knowledge about
the shape of the \runtime distribution when we communicate messages of
size $m_i$ in $k$ rounds. As a matter of fact, we cannot simply shift
the distribution at $m_i$ by some constant factor, and therefore, we
decided to rely on (and to compare) the median values of the
distributions.

Since we measure the \runtime of MPI functions only for a limited
number of message sizes, we compute the factor
$k = \min_{l \in \mathbb{N}} ( l m_i \ge m_j )$, which denotes the
smallest multiple of $m_i$ such that the resulting product is at least
$m_j$. Notice that we explicitly allow $l m_i$ to be larger than
$m_j$, which enables us to check whether sending two messages of size
\SI{1024}{\byte} is faster than sending one message of size
\SI{2000}{\byte}. The \pgmpi framework checks whether the time to
communicate messages of size $m_i$ in $k$ rounds is smaller than the
\runtime for $m_j$. If we find such a violation for a message size
$m_j$, we only report the largest message size $m_i$ (the smallest
factor $k$) for which the violation occurred; otherwise too many
violations would be reported in some cases. It often happens that the
predicted \runtime for $l m_i$ is very similar to the \runtime for
$m_j$. To avoid reporting split-robustness violations for which only
marginal relative \runtime differences have been measured, we use a
5\% tolerance level to verify this guideline. Currently, \pgmpi does
not empirically test whether communicating $k$ messages of size $m_i$
is indeed faster than communication a message of size $m_j$ in
practice. This additional check would require an additional
benchmarking round, and might be added to \pgmpi later.

\subsubsection{Pattern Guidelines}

The verification of pattern guidelines is done similarly to checking
the monotony guideline, except that we now compare two \runtime
distributions of two distinct functions: an MPI function and its
mock-up version. We apply the \testwilcox on the two distributions to
test whether the \runtime distribution of the MPI function is not
significantly shifted to the right of the distribution obtained with
the mock-up version (``to the right'' means larger \runtime).
If this is the case, \pgmpi reports a pattern violation.
Alternatively, the \testkolm~\cite{hollander2014} can be employed, as
it is less sensitive to ties. Overall, both tests led to similar
results in the majority of the considered cases.

\section{Experimental Evaluation and Results}
\label{sec:exp_results}

\begin{table*}[t]
  \centering
  \caption{Overview of parallel machines used in the experiments.}
  \label{tab:machines}
  \begin{scriptsize}
  \begin{tabular}{l@{\hskip .1in}l@{\hskip .1in}l}
    \toprule
    name & hardware & MPI libraries / Compiler \\
    \midrule 
    \machone & 36 $\times$ Dual Opteron 6134 @ \SI{2.3}{\giga\hertz}  & \jupiternecmpi, \jupitermvapich \\
             &  \infiniband QDR MT26428  & \jupiteropenmpi / \gcc 4.4.7 \\
    \machtwo &  \num{2000} $\times$ Dual Xeon E5-2650V2 @ \SI{2.6}{\giga\hertz} & \vscintelmpi  \\
             & \infiniband QDR-80 &  \gcc 4.4.7 \\ 
    \machthree    &  \num{10240} $\times$ IBM PowerA2 @ \SI{1.6}{\giga\hertz} & \fermibgqmpi   \\
             & IBM-BlueGene/Q, 5D Torus interconnect &  IBM XL \\ 
    \bottomrule
  \end{tabular}
  \end{scriptsize}
\end{table*}

We evaluate our proposed \pgmpi
framework\footnote{\url{https://github.com/hunsa/pgmpi}}
experimentally using the hardware and software setup listed in
\tab~\ref{tab:machines}. First, we present a summary of detected
performance-guideline violations for several MPI libraries.  Second,
we demonstrate in two case studies that the knowledge about specific
guideline violations can help tuning and adapting MPI
implementations to parallel systems.

\subsection{Assessing the Guideline Compliance of MPI Libraries}

We used the \pgmpi framework to verify the performance guidelines
listed in \append~\ref{sec:appendix_guidelines} for different MPI
libraries.
On \machone, we evaluated \jupiternecmpi, \jupitermvapich, and
\jupiteropenmpi. The \jupiternecmpi library was delivered by NEC
pre-compiled for our system and we therefore do not know all
internals. The other two libraries, \jupitermvapich and
\jupiteropenmpi, were compiled using the default settings.  On
\machtwo, we recorded guideline violations for the proprietary
\vscintelmpi.

\begin{table*}[t]
  \centering
  \caption{\label{tab:gl_check_overview}Performance-guideline violations of different MPI libraries ($\nmpiruns=10$); violation types: \textbf{m}onotony, \textbf{s}plit-robustness, \textbf{p}attern; message sizes between \SI{1}{\byte} and \SI{100}{\kibi\byte}.}
  \begin{minipage}{.4\linewidth}
    \centering {\scriptsize
      (a)~\machone}\\[.4ex]
  \begin{scriptsize}
\begin{tabular}{ccrrr}
  \hline
\#processes & type & {\tiny{}\jupitermvapich} & {\tiny{}\jupiternecmpi} & {\tiny{}\jupiteropenmpi} \\ 
  \hline
16x1 & m & 7/9 & 6/9 & 7/9 \\ 
   \hline
16x1 & s & 1/9 & 0/9 & 3/9 \\ 
   \hline
16x1 & p & 12/15 & 7/15 & 9/15 \\ 
   \hline
32x16 & m & 5/9 & 4/9 & 4/9 \\ 
   \hline
32x16 & s & 3/9 & 0/9 & 3/9 \\ 
   \hline
32x16 & p & 8/15 & 7/15 & 7/15 \\ 
   \hline
\end{tabular}

  \end{scriptsize}    
  \end{minipage}
  \begin{minipage}{.30\linewidth}
    \centering
    {\scriptsize (b)~\machtwo}\\[.4ex]
  \begin{scriptsize}
\begin{tabular}{ccr}
  \hline
\#processes & type & {\tiny{}Intel MPI 5.0} \\ 
  \hline
16x16 & m & 7/9 \\ 
   \hline
16x16 & s & 6/9 \\ 
   \hline
16x16 & p & 13/15 \\ 
   \hline
64x16 & m & 6/9 \\ 
   \hline
64x16 & s & 7/9 \\ 
   \hline
64x16 & p & 11/15 \\ 
   \hline
\end{tabular}

  \end{scriptsize}    
  \end{minipage}
\end{table*}
\tab~\ref{tab:gl_check_overview} presents an overview of the detected
guideline violations for several MPI libraries on \machone and
\machtwo. For the \glmono and the \glsplit guidelines, the table shows
the number of MPI functions for which violations occurred, \eg, for
\jupitermvapich using \num{16x1} processes, \pgmpi found seven
monotony violations among the nine tested MPI collectives.  For the
pattern guidelines, we verified the 15 guidelines provided in
\append~\ref{sec:appendix_guidelines}. If a guideline violation is
found for any message size of a particular MPI function, we say that
this particular guideline is unsatisfied, \ie, a violation is only
counted once across all message sizes. We can observe in
\tab~\ref{tab:gl_check_overview} that the \glmono and the \glsplit
guidelines are violated by approximately 50\% of the collectives. The
table also reveals that more than 40\% of the examined pattern
guidelines were violated. The guideline violations occurred across
different numbers of processes, message sizes, libraries, and
machines. We therefore contend that there is a large potential for
optimization of the individual MPI libraries on these machines.

\begin{table*}[t]
  \centering
  \caption{\label{tab:gl_patterns_512}Pattern guideline violations of different MPI libraries with \num{32x16} processes on \machone, $\nmpiruns=10$; message sizes between \SI{1}{\byte} and \SI{100}{\kibi\byte}.}
  \begin{scriptsize}
\begin{tabular}{lccc}
  \hline
guideline & {\tiny{}\jupitermvapich} & {\tiny{}\jupiternecmpi} & {\tiny{}\jupiteropenmpi} \\ 
  \hline
\mpiallgather \guidelt Allreduce & \scalebox{0.7}{\textbullet} &  &  \\ 
   \hline
\mpiallgather \guidelt Alltoall &  &  &  \\ 
   \hline
\mpiallgather \guidelt Gather$+$Bcast &  & \scalebox{0.7}{\textbullet} &  \\ 
   \hline
\mpiallreduce \guidelt Reduce$+$Bcast & \scalebox{0.7}{\textbullet} & \scalebox{0.7}{\textbullet} & \scalebox{0.7}{\textbullet} \\ 
   \hline
\mpiallreduce \guidelt Reduce\_scatter\_block$+$Allgather &  &  & \scalebox{0.7}{\textbullet} \\ 
   \hline
\mpibcast \guidelt Scatter$+$Allgather &  & \scalebox{0.7}{\textbullet} & \scalebox{0.7}{\textbullet} \\ 
   \hline
\mpigather \guidelt Allgather &  &  &  \\ 
   \hline
\mpigather \guidelt Reduce & \scalebox{0.7}{\textbullet} &  &  \\ 
   \hline
\mpireducescatterblock \guidelt Reduce+Scatter & \scalebox{0.7}{\textbullet} & \scalebox{0.7}{\textbullet} & \scalebox{0.7}{\textbullet} \\ 
   \hline
\mpireducescatter \guidelt Allreduce & \scalebox{0.7}{\textbullet} & \scalebox{0.7}{\textbullet} & \scalebox{0.7}{\textbullet} \\ 
   \hline
\mpireducescatter \guidelt Reduce$+$Scatterv &  & \scalebox{0.7}{\textbullet} &  \\ 
   \hline
\mpireduce \guidelt Allreduce & \scalebox{0.7}{\textbullet} &  & \scalebox{0.7}{\textbullet} \\ 
   \hline
\mpireduce \guidelt Reduce\_scatter\_block$+$Gather & \scalebox{0.7}{\textbullet} &  &  \\ 
   \hline
\mpiscan \guidelt Exscan$+$Reduce\_local &  & \scalebox{0.7}{\textbullet} & \scalebox{0.7}{\textbullet} \\ 
   \hline
\mpiscatter \guidelt Bcast & \scalebox{0.7}{\textbullet} &  &  \\ 
   \hline
\end{tabular}

  \end{scriptsize}    
\end{table*}
\tab~\ref{tab:gl_patterns_512} compares the detected pattern
violations for the three libraries on \machone. Except for two
guidelines, we found violations for all other \mbox{pattern}
guidelines in at least one MPI library. The experimental results
clearly suggest that Reduce-like functions should be improved and
tuned in all MPI libraries, which are: \mpiallreduce, \mpireduce,
\mpireducescatter, and \mpireducescatterblock.

\begin{table*}[t]
  \centering
  \caption{\label{tab:mvapich_case1}Performance-guideline violations of \jupitermvapich using \num{32x16} processes on \machone ($\nmpiruns=10$); violation types: \textbf{m}onotony, \textbf{s}plit-robustness, \textbf{p}attern.}
  \begin{scriptsize}
\begin{tabular}{clacacacacacacacacacaca}
  \hline
type & function & \begin{sideways}\num{1}\end{sideways} & \begin{sideways}\num{2}\end{sideways} & \begin{sideways}\num{4}\end{sideways} & \begin{sideways}\num{8}\end{sideways} & \begin{sideways}\num{16}\end{sideways} & \begin{sideways}\num{32}\end{sideways} & \begin{sideways}\num{64}\end{sideways} & \begin{sideways}\num{100}\end{sideways} & \begin{sideways}\num{128}\end{sideways} & \begin{sideways}\num{256}\end{sideways} & \begin{sideways}\num{512}\end{sideways} & \begin{sideways}\num{1024}\end{sideways} & \begin{sideways}\num{1500}\end{sideways} & \begin{sideways}\num{2048}\end{sideways} & \begin{sideways}\num{4096}\end{sideways} & \begin{sideways}\num{5000}\end{sideways} & \begin{sideways}\num{8192}\end{sideways} & \begin{sideways}\num{10000}\end{sideways} & \begin{sideways}\num{16384}\end{sideways} & \begin{sideways}\num{32768}\end{sideways} & \begin{sideways}\num{102400}\end{sideways} \\ 
  \hline
m & \mpiallgather &  &  & \scalebox{0.7}{\textbullet} &  &  & \scalebox{0.7}{\textbullet} &  &  &  &  &  &  &  &  &  &  &  &  &  &  &  \\ 
   \hline
m & \mpiallreduce &  &  &  &  &  &  & \scalebox{0.7}{\textbullet} &  &  &  &  &  &  &  &  &  &  &  & \scalebox{0.7}{\textbullet} &  &  \\ 
   \hline
m & \mpigather &  &  &  &  &  &  &  &  &  &  &  &  &  &  &  &  &  &  & \scalebox{0.7}{\textbullet} &  & \scalebox{0.7}{\textbullet} \\ 
   \hline
m & \mpireduce &  &  &  &  &  &  & \scalebox{0.7}{\textbullet} &  &  &  &  &  &  &  &  &  &  &  & \scalebox{0.7}{\textbullet} &  &  \\ 
   \hline
m & \mpiscatter &  & \scalebox{0.7}{\textbullet} &  & \scalebox{0.7}{\textbullet} &  & \scalebox{0.7}{\textbullet} &  &  &  &  &  &  &  &  &  &  &  &  &  &  &  \\ 
   \hline
s & \mpigather &  &  &  &  &  &  &  &  &  &  &  &  &  & \scalebox{0.7}{\textbullet} & \scalebox{0.7}{\textbullet} & \scalebox{0.7}{\textbullet} & \scalebox{0.7}{\textbullet} & \scalebox{0.7}{\textbullet} &  & \scalebox{0.7}{\textbullet} &  \\ 
   \hline
s & \mpireduce &  &  &  &  &  &  &  &  &  &  &  &  &  &  &  &  &  & \scalebox{0.7}{\textbullet} &  &  &  \\ 
   \hline
s & \mpireducescatterblock &  &  &  &  &  &  &  &  &  &  &  & \scalebox{0.7}{\textbullet} & \scalebox{0.7}{\textbullet} &  &  &  &  &  &  &  &  \\ 
   \hline
p & \mpiallgather \guidelt Allreduce & \scalebox{0.7}{\textbullet} & \scalebox{0.7}{\textbullet} &  &  &  &  &  &  &  &  &  &  &  &  &  &  &  &  &  &  &  \\ 
   \hline
p & \mpiallreduce \guidelt Reduce$+$Bcast & \scalebox{0.7}{\textbullet} & \scalebox{0.7}{\textbullet} & \scalebox{0.7}{\textbullet} & \scalebox{0.7}{\textbullet} & \scalebox{0.7}{\textbullet} & \scalebox{0.7}{\textbullet} & \scalebox{0.7}{\textbullet} & \scalebox{0.7}{\textbullet} & \scalebox{0.7}{\textbullet} & \scalebox{0.7}{\textbullet} & \scalebox{0.7}{\textbullet} & \scalebox{0.7}{\textbullet} & \scalebox{0.7}{\textbullet} & \scalebox{0.7}{\textbullet} &  &  &  &  &  &  &  \\ 
   \hline
p & \mpigather \guidelt Reduce & \scalebox{0.7}{\textbullet} & \scalebox{0.7}{\textbullet} & \scalebox{0.7}{\textbullet} &  &  &  &  &  &  &  &  &  &  &  &  &  &  &  &  &  &  \\ 
   \hline
p & \mpireducescatterblock \guidelt Reduce+Scatter & \scalebox{0.7}{\textbullet} & \scalebox{0.7}{\textbullet} &  &  &  &  &  &  &  &  &  & \scalebox{0.7}{\textbullet} & \scalebox{0.7}{\textbullet} &  &  &  &  &  &  &  &  \\ 
   \hline
p & \mpireducescatter \guidelt Allreduce & \scalebox{0.7}{\textbullet} & \scalebox{0.7}{\textbullet} &  &  &  &  & \scalebox{0.7}{\textbullet} & \scalebox{0.7}{\textbullet} & \scalebox{0.7}{\textbullet} & \scalebox{0.7}{\textbullet} & \scalebox{0.7}{\textbullet} & \scalebox{0.7}{\textbullet} & \scalebox{0.7}{\textbullet} & \scalebox{0.7}{\textbullet} & \scalebox{0.7}{\textbullet} & \scalebox{0.7}{\textbullet} & \scalebox{0.7}{\textbullet} & \scalebox{0.7}{\textbullet} & \scalebox{0.7}{\textbullet} & \scalebox{0.7}{\textbullet} & \scalebox{0.7}{\textbullet} \\ 
   \hline
p & \mpireduce \guidelt Allreduce &  &  &  &  &  &  &  &  &  &  &  &  &  &  &  &  &  & \scalebox{0.7}{\textbullet} & \scalebox{0.7}{\textbullet} & \scalebox{0.7}{\textbullet} & \scalebox{0.7}{\textbullet} \\ 
   \hline
p & \mpireduce \guidelt Reduce\_scatter\_block$+$Gather &  &  &  &  &  &  &  &  &  &  &  &  &  &  &  &  &  &  & \scalebox{0.7}{\textbullet} & \scalebox{0.7}{\textbullet} & \scalebox{0.7}{\textbullet} \\ 
   \hline
p & \mpiscatter \guidelt Bcast & \scalebox{0.7}{\textbullet} & \scalebox{0.7}{\textbullet} & \scalebox{0.7}{\textbullet} &  &  &  &  &  &  &  &  &  &  &  &  &  &  &  &  &  &  \\ 
   \hline
\end{tabular}

  \end{scriptsize}
\end{table*}

A detailed view on the detected guideline violations for
\jupitermvapich on \machone is given in \tab~\ref{tab:mvapich_case1}.
For this MPI library, we observe a couple of monotony violations. For
short message sizes ($<\SI{32}{\byte}$), the absolute difference in
\runtimes is very small, and thus, fixing these cases has low
priority.  Monotony violations occur for larger messages when the
message size is not a power of two (\eg, between \SI{10000}{\byte} and
\SI{16384}{\byte}). These cases could be investigated in more detail,
as padding up the message to the next power of two could be an
option. For \glsplit guidelines we can see potential for improvement
only for larger message sizes.  When analyzing the pattern guidelines,
two cases stand out: \mpiallreduce is slower than the emulating
function using Reduce and Bcast for message sizes up to
\SI{2}{\kibi\byte} and \mpireducescatter exposes a performance
degradation compared to Allreduce for almost all message sizes.

As it is impossible for library developers to provide suitable
parameters for each individual installation, checking the compliance
to performance guidelines can be seen as indicators for programmers
and administrators, how to tune MPI libraries.  Often, specific MPI
libraries already provide efficient algorithms, and violations would
not occur if the right algorithm were enabled for a specific case. We
therefore show in the next section how violations can guide us to find
more suitable algorithms and implementations for collective calls on a
specific machine.

\subsection{Case Study 1: \mpigather \guidelt \mpiallgather, \mvapich}

\begin{figure*}[t]
  \centering
  \subfloat[][With Violations]{
    \includegraphics[width=.45\linewidth]{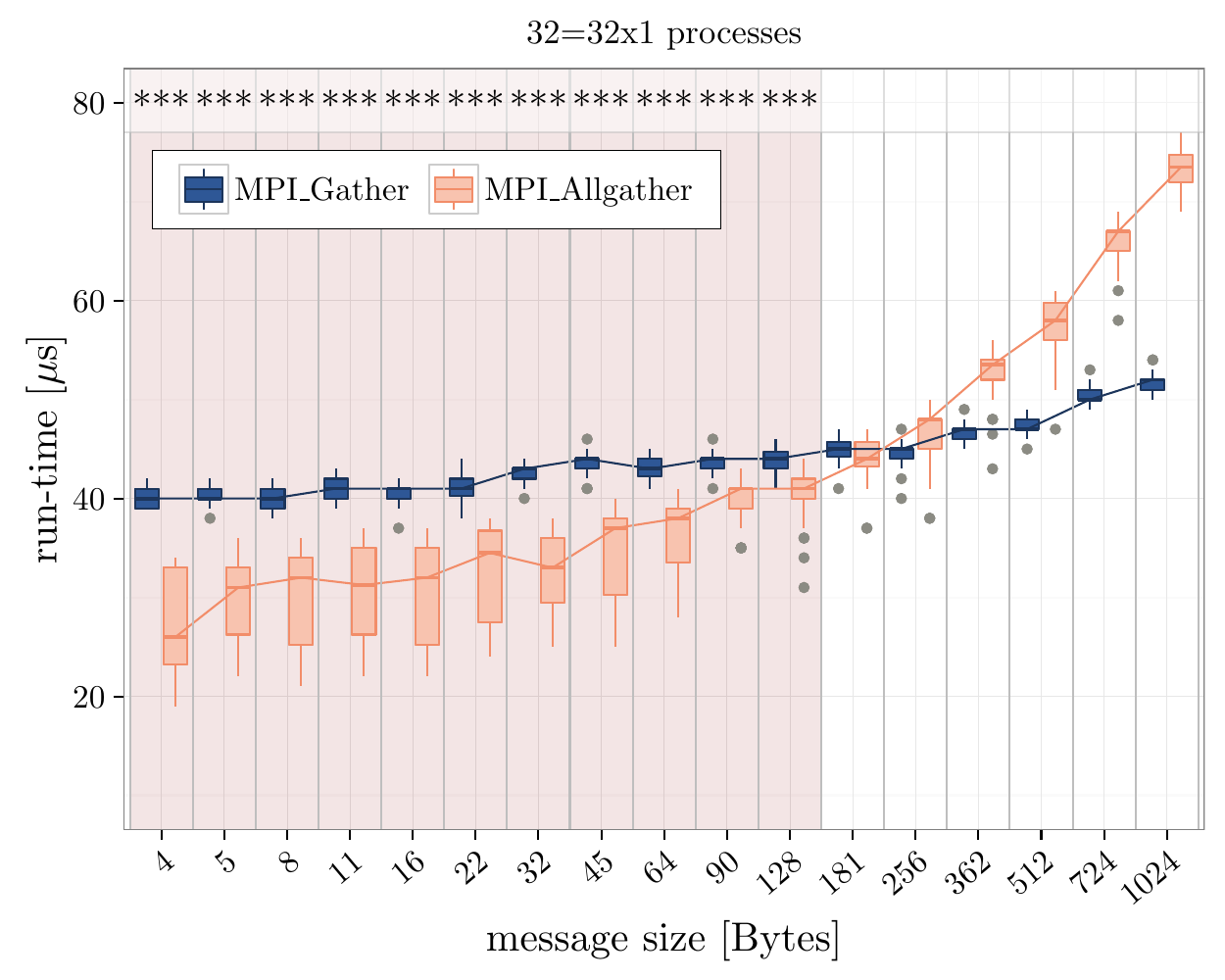}
    \label{fig:mvapich_gather_vs_allgatherA}
  }
  \hfill
  \subfloat[][No Violations]{
    \includegraphics[width=.45\linewidth]{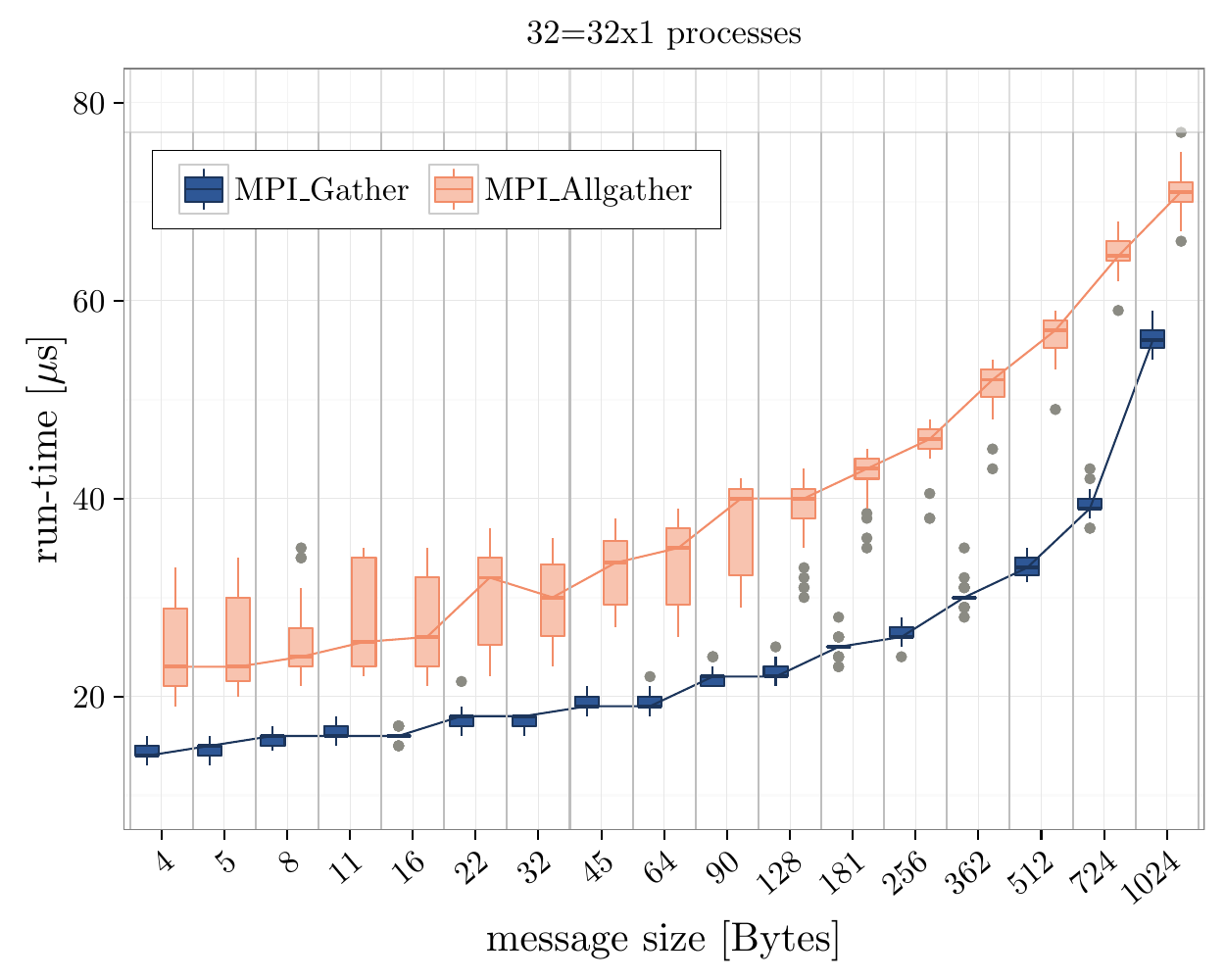}
    \label{fig:mvapich_gather_vs_allgatherB}
    }
    \caption{\label{fig:mvapich_gather_vs_allgather} Verification of
      \mpigather \guidelt \mpiallgather (a)~before and (b)~after
      changing the Gather implementation (\jupitermvapich, \machone,
      $\nmpiruns=\num{30}$, $\nrep=\num{1000}$).}
\end{figure*}

We consider the violation of this performance guideline that was
detected using \num{32x1} processes and \jupitermvapich on \machone
and is shown in \Fig~\ref{fig:mvapich_gather_vs_allgatherA}.  When the
\testwilcox reports a violation for a particular message size, we mark
this case in the figure with a red background and add asterisks to
show the statistical significance.
Here, executing \mpigather using \num{32x1} processes (one process per
compute node) is slower than performing a Gather using \mpiallgather.
Calling \mpigather in the default installation of
\jupitermvapich will use the internal function
\code{MPIR\_Gather\_intra} for the first \num{14} invocations and then
switch to \code{MPIR\_Gather\_MV2\_Direct} for subsequent calls. The
direct implementation of Gather performs $(\np-1)$ \mpiirecv{}s on the
root process and an \mpisend on the other processes. We can set the
environment variable \code{MV2\_USE\_DIRECT\_GATHER=0} to force
\mvapich to use \code{MPIR\_Gather\_intra} only. The intra-version on
our machine uses a binomial tree algorithm to implement Gather, and
forcing this algorithm fixes the violation (\cf
\Fig~\ref{fig:mvapich_gather_vs_allgatherB}).
Let us check that the algorithmic change for small message sizes is
indeed favorable. We use the Hockney model for \mpireduce given by
Pjesivac~\etal~\cite{Pjesivac-GrbovicABFGD07}, but omit the computational
term. As the direct algorithm issues
$(p-1)$ receive operations, we obtain a \runtime for small message
sizes (we neglect the bandwidth term) of about
\SI{52.7}{\micro\second}, for a network latency of roughly
\SI{1.7}{\micro\second}.  If we use a binomial tree algorithm instead,
the latency cost grows only logarithmically in the number of
processes, \ie,
$\log{32}\cdot\SI{1.7}{\micro\second}=\SI{8.5}{\micro\second}$.  Even
though our estimation does not perfectly match the experimental data,
it explains why the binomial tree algorithm performs better.

\subsection{Case Study 2: \mpireduce  \guidelt \mpiallreduce, \openmpi}

\begin{figure*}[t!]
  \centering
  \subfloat[][With Violations]{
    \includegraphics[width=.45\linewidth]{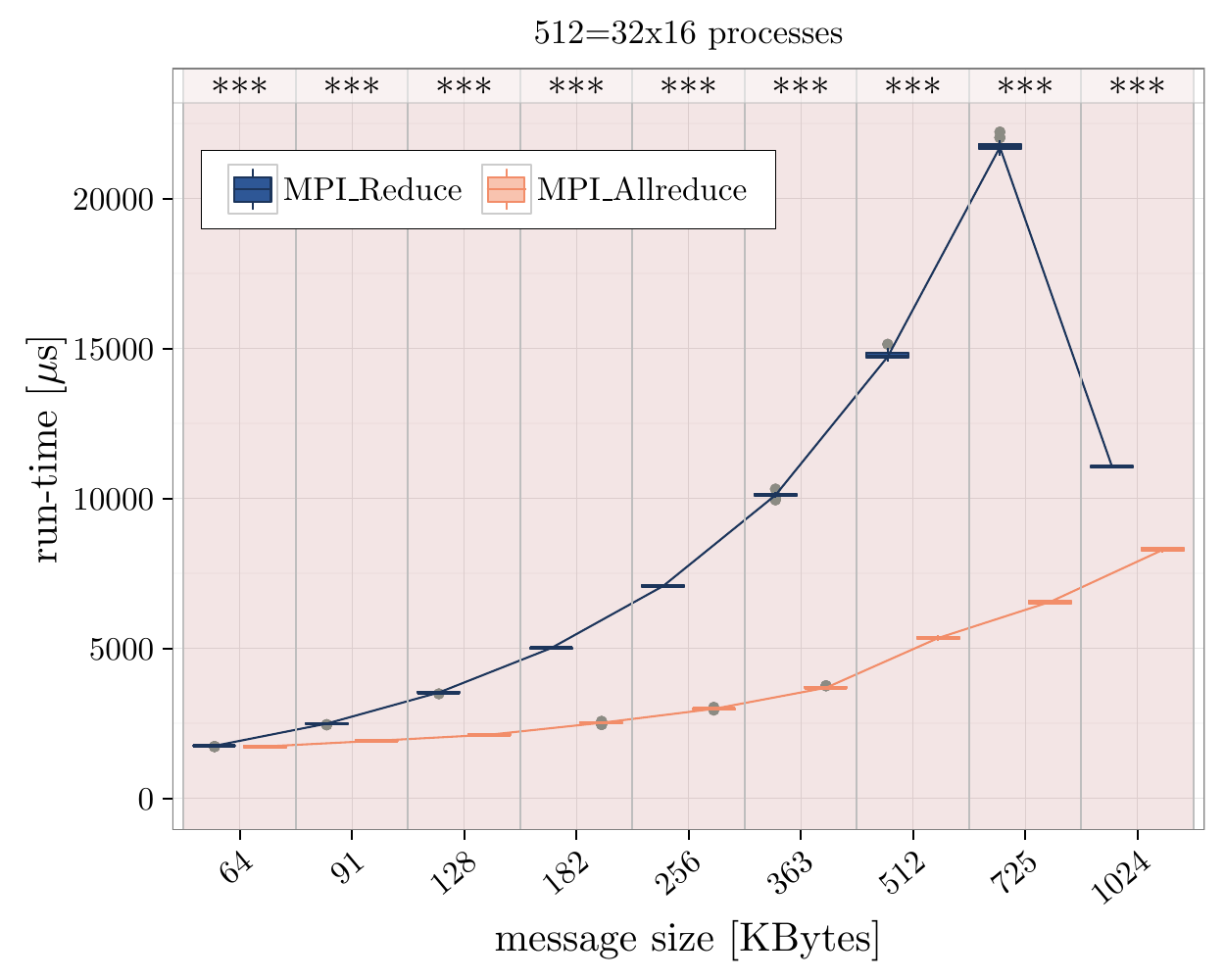}
    \label{fig:openmpi_reduce_vs_allreduceA}
  }
  \hfill
  \subfloat[][No Violations]{
    \includegraphics[width=.45\linewidth]{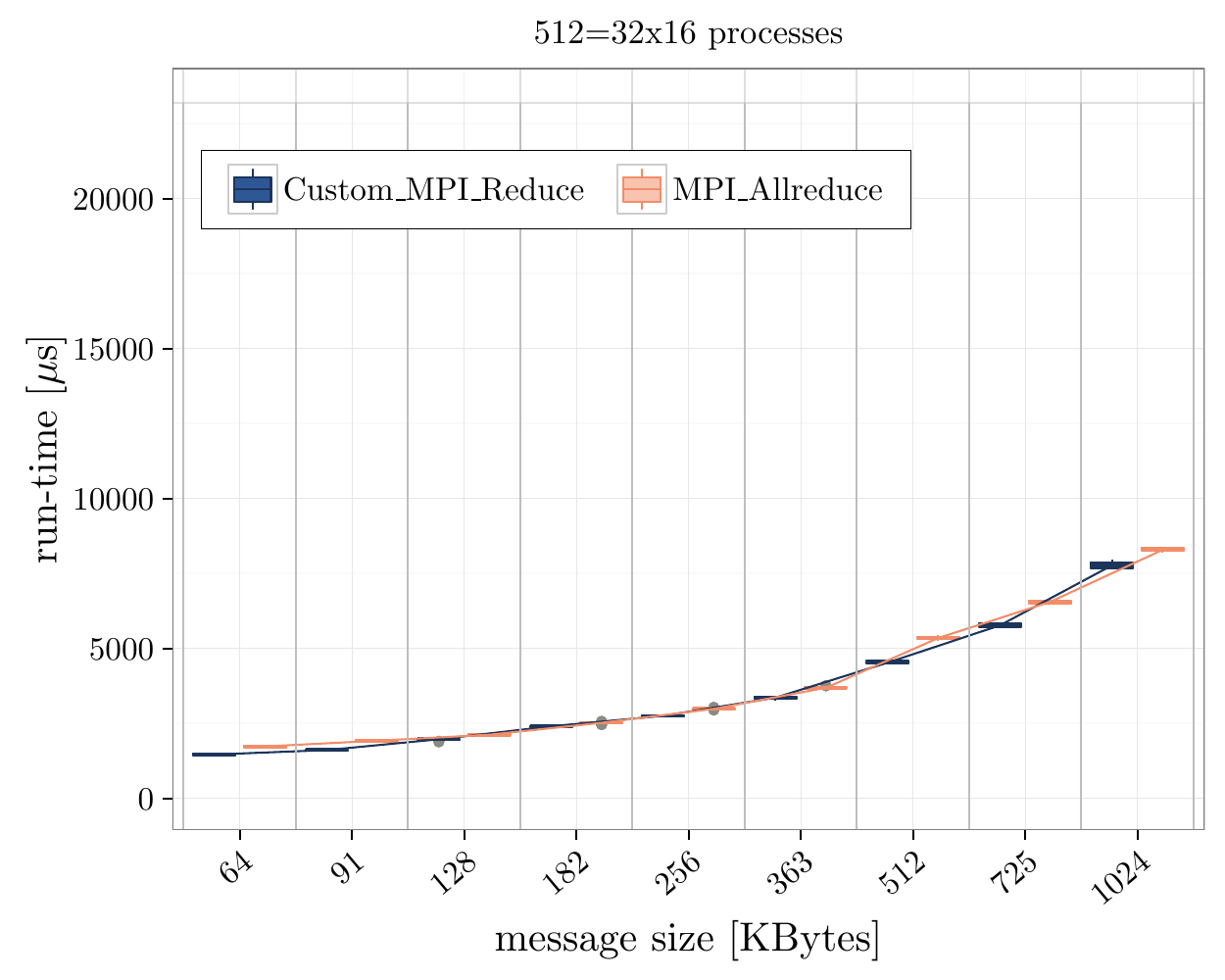}    
    \label{fig:openmpi_reduce_vs_allreduceB}
  }
  \caption{\label{fig:openmpi_reduce_vs_allreduce} Verification of
    \mpireduce \guidelt \mpiallreduce with (a) original and (b) new
    Reduce implementation (\jupiteropenmpi, \machone,
    $\nmpiruns=\num{5}$, $\nrep=\num{100}$).}
\end{figure*}

In the second case study, we consider the guideline violations that
occurred for \mpireduce using \jupiteropenmpi on the \machone
system. Here, in contrast to the first case study, violations have
only been measured for larger message sizes
($>\SI[exponent-base = 2]{e16}{\byte}$), but for various numbers of
processes: \num{16x1}, \num{32x1}, \num{16x16}, and \num{32x16}.
\fig~\ref{fig:openmpi_reduce_vs_allreduce} limits the view to message
sizes for which violations were detected.
Since \openmpi is highly configurable via the \texttt{MCA} parameters,
we have tried to find parameter settings for \mpireduce, such that
executing the latter would be faster than executing \mpiallreduce. We
have tried various segment sizes and fan-outs (where the parameters
were applicable). Unfortunately, we failed to tune the parameters in
such a way that the violations would disappear on our machine. For
that reason, we implemented our own Reduce algorithm, which is based
on the \mpiallreduce algorithm found in \jupiteropenmpi. Here,
\mpiallreduce is implemented using a Reduce-scatter followed by an
Allgatherv on a ring of processes~\cite{PatarasukY09}. We modified
this algorithm to become an \mpireduce by replacing the final
Allgatherv by a Gatherv to the root. The Gatherv was realized using a
direct Irecv/Send scheme. In the MPI semantics of Reduce, only the
root process has a receive buffer. We therefore need to allocate
additional buffer space to send and receive data segments in the
Reduce-scatter phase. We found that executing \texttt{malloc} in each
Reduce call has a severe impact on the performance of Reduce. To
overcome this problem, we allocate a temporary buffer outside of
Reduce but accessible to the Reduce implementation. This modification
helped us to significantly speed up the \runtime, and made this Reduce
implementation a suitable candidate to be included in the \openmpi
library.
In sum, our Reduce implementation avoids violations for larger
messages sizes, as shown in
\fig~\ref{fig:openmpi_reduce_vs_allreduceB}.

\section{Conclusions}
\label{sec:conclusions}

The experimental verification of performance guidelines is an
orthogonal approach to traditional MPI library tuning. It allows to
find performance degradations of MPI functions, which would be hidden
otherwise. For example, it is possible to optimize several existing
implementations of \mpigather, but even the fastest of these Gather
algorithms might be slower than the call to \mpiallgather.

We have introduced the \pgmpi framework to verify 
self-consistent performance guidelines of MPI functions. Currently,
the framework supports blocking MPI collective communication
operations, but it can be extended to support MPI point-to-point
communication operations and derived datatypes. We have evaluated 17
different guidelines for collective communication operations for
several MPI libraries such as \mvapich and \openmpi. The experimental
results reveal that none of the libraries was well adapted to our
parallel machines, which might not be surprising. However, by using
\pgmpi we were able to pinpoint exactly which MPI functions should be
tuned and which message sizes should be considered. Thus, \pgmpi is a
useful tool for MPI developers and system administrators to easily
spot tuning potentials.

\section*{Acknowledgements}

We acknowledge PRACE for awarding us access to resource FERMI based in
Italy at CINECA, Bologna.

\IEEEtriggeratref{9}
\bibliographystyle{IEEEtran}
\bibliography{mpi_guidelines}

\onecolumn
\appendices

\section{Self-Consistent Performance Guidelines in \pgmpi}
\label{sec:appendix_guidelines}

\renewcommand{\theequation}{GL\arabic{equation}}
\setcounter{equation}{0}

The guidelines are formulated for a variable communication volume~$n,\, n \ge 0$
and fixed number of processes~$p,\, p \ge 1$, which is omitted.

\paragraph*{Monotony Guideline}

\begin{align}
\texttt{MPI}\_A(n) \guidelt \texttt{MPI}\_A(n + k) \quad , \qquad k \ge 0 \label{gl:monotony}
\end{align}

\paragraph*{Split-Robustness Guideline}

\begin{align}
\texttt{MPI}\_A(n) \guidelt k\, \texttt{MPI}\_A(n/k) \quad , \qquad k \ge 1 \label{gl:split}
\end{align}

\paragraph*{Pattern Guidelines}

\begin{align}
	\mpigather(n) & \guidelt \mpiallgather(n) \label{gl:gather_lt_allgather} \\
	\mpigather(n) & \guidelt \mpireduce(n) \\
	\mpiallgather(n) & \guidelt \mpialltoall(n) \\
	\mpiallgather(n) & \guidelt \mpiallreduce(n) \\
	\mpiscatter(n) & \guidelt \mpibcast(n) \\
	\mpireduce(n) & \guidelt \mpiallreduce(n) \label{gl:reduce_lt_allreduce} \\
	\mpireducescatter(n) & \guidelt \mpiallreduce(n) \\
	\mpibcast(n) & \guidelt \mpiscatter(n) + \mpiallgather(n) \label{gl:bcast_lt_scatter_allgather}\\
	\mpiallgather(n) & \guidelt \mpigather(n) + \mpibcast(n) \\
	\mpiallreduce(n) & \guidelt \mpireduce(n) + \mpibcast(n) \\
	\mpiallreduce(n) & \guidelt \mpireducescatterblock(n) \nonumber \\
                         & \quad + \mpiallgather(n) \\
	\mpireduce(n) & \guidelt \mpireducescatterblock(n) \nonumber \\ 
                      & \quad + \mpigather(n) \\
	\mpireducescatterblock(n) & \guidelt \mpireduce(n) + \mpiscatter(n) \\
	\mpiscan(n) & \guidelt \mpiexscan(n) + \mpireducelocal(n) \\
 	\mpireducescatter(n) & \guidelt \mpireduce(n) + \mpiscatterv(n) 
\end{align}

\clearpage
\section{Further Experimental Results}

\subsection{\machone}

\begin{table*}[!h]
  \centering
  \caption{Performance-guideline violations of \jupitermvapich using \num{16x1} processes on \machone ($\nmpiruns=10$); \newline violation types: \textbf{m}onotony, \textbf{s}plit-robustness, \textbf{p}attern.}
  \begin{small}
\begin{tabular}{clacacacacacacacacacac}
  \hline
type & function & \begin{sideways}\num{1}\end{sideways} & \begin{sideways}\num{2}\end{sideways} & \begin{sideways}\num{4}\end{sideways} & \begin{sideways}\num{8}\end{sideways} & \begin{sideways}\num{16}\end{sideways} & \begin{sideways}\num{32}\end{sideways} & \begin{sideways}\num{64}\end{sideways} & \begin{sideways}\num{100}\end{sideways} & \begin{sideways}\num{128}\end{sideways} & \begin{sideways}\num{256}\end{sideways} & \begin{sideways}\num{512}\end{sideways} & \begin{sideways}\num{1024}\end{sideways} & \begin{sideways}\num{1500}\end{sideways} & \begin{sideways}\num{2048}\end{sideways} & \begin{sideways}\num{4096}\end{sideways} & \begin{sideways}\num{5000}\end{sideways} & \begin{sideways}\num{8192}\end{sideways} & \begin{sideways}\num{10000}\end{sideways} & \begin{sideways}\num{16384}\end{sideways} & \begin{sideways}\num{32768}\end{sideways} \\ 
  \hline
m & \mpibcast &  &  & \scalebox{0.7}{\textbullet} &  &  &  &  &  &  &  &  &  &  &  &  &  &  &  &  &  \\ 
   \hline
m & \mpigather &  &  &  &  &  &  &  &  & \scalebox{0.7}{\textbullet} &  &  &  &  &  &  &  &  &  &  &  \\ 
   \hline
m & \mpiscatter &  & \scalebox{0.7}{\textbullet} &  &  &  &  & \scalebox{0.7}{\textbullet} &  &  &  & \scalebox{0.7}{\textbullet} &  &  &  &  &  &  &  &  &  \\ 
   \hline
p & \mpiallgather \guidelt Alltoall &  &  &  &  &  &  &  &  &  &  &  &  &  &  &  &  & \scalebox{0.7}{\textbullet} & \scalebox{0.7}{\textbullet} & \scalebox{0.7}{\textbullet} &  \\ 
   \hline
p & \mpibcast \guidelt Scatter$+$Allgather &  &  &  &  &  &  &  &  &  &  &  &  &  &  &  &  & \scalebox{0.7}{\textbullet} & \scalebox{0.7}{\textbullet} & \scalebox{0.7}{\textbullet} & \scalebox{0.7}{\textbullet} \\ 
   \hline
p & \mpireduce \guidelt Allreduce &  &  &  &  &  &  &  &  &  &  &  &  &  &  &  &  & \scalebox{0.7}{\textbullet} & \scalebox{0.7}{\textbullet} & \scalebox{0.7}{\textbullet} & \scalebox{0.7}{\textbullet} \\ 
   \hline
\end{tabular}

  \end{small}
\end{table*}

\FloatBarrier

\subsection{\machtwo}

\begin{table}[!h]
  \centering
  \caption{\label{tab:vsc_case1}Performance-guideline violations of \vscintelmpi using \num{16x1} processes on  \machtwo ($\nmpiruns=30$); violation types: \textbf{m}onotony, \textbf{s}plit-robustness, \textbf{p}attern.}
  \begin{scriptsize}
\begin{tabular}{clacacacacacacacacacacac}
  \hline
type & function & \begin{sideways}\num{1}\end{sideways} & \begin{sideways}\num{2}\end{sideways} & \begin{sideways}\num{4}\end{sideways} & \begin{sideways}\num{8}\end{sideways} & \begin{sideways}\num{16}\end{sideways} & \begin{sideways}\num{32}\end{sideways} & \begin{sideways}\num{64}\end{sideways} & \begin{sideways}\num{100}\end{sideways} & \begin{sideways}\num{128}\end{sideways} & \begin{sideways}\num{256}\end{sideways} & \begin{sideways}\num{512}\end{sideways} & \begin{sideways}\num{1024}\end{sideways} & \begin{sideways}\num{1500}\end{sideways} & \begin{sideways}\num{2048}\end{sideways} & \begin{sideways}\num{4096}\end{sideways} & \begin{sideways}\num{5000}\end{sideways} & \begin{sideways}\num{8192}\end{sideways} & \begin{sideways}\num{10000}\end{sideways} & \begin{sideways}\num{16384}\end{sideways} & \begin{sideways}\num{32768}\end{sideways} & \begin{sideways}\num{102400}\end{sideways} & \begin{sideways}\num{1048576}\end{sideways} \\ 
  \hline
m & \mpiallreduce &  &  & \scalebox{0.7}{\textbullet} &  &  &  & \scalebox{0.7}{\textbullet} &  &  &  &  & \scalebox{0.7}{\textbullet} &  &  & \scalebox{0.7}{\textbullet} &  &  &  &  &  &  &  \\ 
   \hline
m & \mpibcast &  &  &  &  &  &  &  &  &  &  &  &  &  &  &  &  &  & \scalebox{0.7}{\textbullet} &  &  &  &  \\ 
   \hline
m & \mpireduce &  & \scalebox{0.7}{\textbullet} & \scalebox{0.7}{\textbullet} & \scalebox{0.7}{\textbullet} &  &  &  &  &  &  &  &  &  &  &  &  &  &  &  &  &  &  \\ 
   \hline
s & \mpiallreduce &  &  &  &  &  &  &  &  &  &  & \scalebox{0.7}{\textbullet} &  &  &  &  &  &  &  &  &  &  &  \\ 
   \hline
p & \mpiallgather \guidelt Alltoall &  &  &  &  &  &  &  &  &  &  & \scalebox{0.7}{\textbullet} & \scalebox{0.7}{\textbullet} & \scalebox{0.7}{\textbullet} & \scalebox{0.7}{\textbullet} & \scalebox{0.7}{\textbullet} & \scalebox{0.7}{\textbullet} & \scalebox{0.7}{\textbullet} & \scalebox{0.7}{\textbullet} & \scalebox{0.7}{\textbullet} & \scalebox{0.7}{\textbullet} & \scalebox{0.7}{\textbullet} & \scalebox{0.7}{\textbullet} \\ 
   \hline
p & \mpiallreduce \guidelt Reduce$+$Bcast &  &  &  &  &  &  &  &  &  &  & \scalebox{0.7}{\textbullet} &  &  &  &  &  &  &  &  &  &  &  \\ 
   \hline
p & \mpiallreduce \guidelt Reduce\_scatter\_block$+$Allgather &  &  &  &  &  &  &  &  &  &  & \scalebox{0.7}{\textbullet} &  &  & \scalebox{0.7}{\textbullet} &  &  &  &  &  &  &  &  \\ 
   \hline
p & \mpibcast \guidelt Scatter$+$Allgather &  &  &  &  &  &  &  &  &  &  &  &  &  &  & \scalebox{0.7}{\textbullet} & \scalebox{0.7}{\textbullet} & \scalebox{0.7}{\textbullet} & \scalebox{0.7}{\textbullet} & \scalebox{0.7}{\textbullet} & \scalebox{0.7}{\textbullet} &  & \scalebox{0.7}{\textbullet} \\ 
   \hline
p & \mpireduce \guidelt Allreduce & \scalebox{0.7}{\textbullet} & \scalebox{0.7}{\textbullet} & \scalebox{0.7}{\textbullet} &  &  &  &  &  &  &  &  &  &  &  &  &  &  & \scalebox{0.7}{\textbullet} &  & \scalebox{0.7}{\textbullet} & \scalebox{0.7}{\textbullet} &  \\ 
   \hline
p & \mpireduce \guidelt Reduce\_scatter\_block$+$Gather &  &  &  &  &  &  &  &  &  &  &  &  &  &  &  &  & \scalebox{0.7}{\textbullet} &  & \scalebox{0.7}{\textbullet} & \scalebox{0.7}{\textbullet} & \scalebox{0.7}{\textbullet} & \scalebox{0.7}{\textbullet} \\ 
   \hline
p & \mpireducescatter \guidelt Allreduce &  & \scalebox{0.7}{\textbullet} & \scalebox{0.7}{\textbullet} & \scalebox{0.7}{\textbullet} &  &  &  &  &  &  &  &  &  &  &  &  &  &  &  &  &  &  \\ 
   \hline
p & \mpiscan \guidelt Exscan$+$Reduce\_local &  &  &  &  &  &  &  &  &  &  &  &  &  &  &  &  &  &  & \scalebox{0.7}{\textbullet} & \scalebox{0.7}{\textbullet} &  &  \\ 
   \hline
\end{tabular}

  \end{scriptsize}
\end{table}

\begin{table}[!h]
  \centering
  \caption{\label{tab:vsc_case2}Performance-guideline violations of \vscintelmpi using \num{16x16} processes on \machtwo ($\nmpiruns=10$);  violation types: \textbf{m}onotony, \textbf{s}plit-robustness, \textbf{p}attern.}
  \begin{scriptsize}
\begin{tabular}{clacacacacacacacacacacac}
  \hline
type & function & \begin{sideways}\num{1}\end{sideways} & \begin{sideways}\num{2}\end{sideways} & \begin{sideways}\num{4}\end{sideways} & \begin{sideways}\num{8}\end{sideways} & \begin{sideways}\num{16}\end{sideways} & \begin{sideways}\num{32}\end{sideways} & \begin{sideways}\num{64}\end{sideways} & \begin{sideways}\num{100}\end{sideways} & \begin{sideways}\num{128}\end{sideways} & \begin{sideways}\num{256}\end{sideways} & \begin{sideways}\num{512}\end{sideways} & \begin{sideways}\num{1024}\end{sideways} & \begin{sideways}\num{1500}\end{sideways} & \begin{sideways}\num{2048}\end{sideways} & \begin{sideways}\num{4096}\end{sideways} & \begin{sideways}\num{5000}\end{sideways} & \begin{sideways}\num{8192}\end{sideways} & \begin{sideways}\num{10000}\end{sideways} & \begin{sideways}\num{16384}\end{sideways} & \begin{sideways}\num{32768}\end{sideways} & \begin{sideways}\num{102400}\end{sideways} & \begin{sideways}\num{1048576}\end{sideways} \\ 
  \hline
m & \mpiallgather &  &  &  & \scalebox{0.7}{\textbullet} &  &  &  &  &  &  &  &  &  &  &  &  &  &  &  &  &  &  \\ 
   \hline
m & \mpiallreduce &  &  &  &  & \scalebox{0.7}{\textbullet} &  &  &  &  &  &  & \scalebox{0.7}{\textbullet} &  & \scalebox{0.7}{\textbullet} & \scalebox{0.7}{\textbullet} &  &  &  &  &  &  &  \\ 
   \hline
m & \mpibcast &  &  &  &  &  &  &  &  &  &  &  &  &  &  &  &  &  & \scalebox{0.7}{\textbullet} &  &  & \scalebox{0.7}{\textbullet} &  \\ 
   \hline
m & \mpigather &  &  &  &  &  &  &  &  & \scalebox{0.7}{\textbullet} &  &  &  &  &  &  &  &  &  &  &  &  &  \\ 
   \hline
m & \mpireduce &  &  & \scalebox{0.7}{\textbullet} & \scalebox{0.7}{\textbullet} &  &  &  &  &  &  &  &  &  &  &  &  & \scalebox{0.7}{\textbullet} &  &  &  &  &  \\ 
   \hline
m & \mpiscan &  & \scalebox{0.7}{\textbullet} &  &  &  &  &  &  &  &  &  &  &  &  &  &  &  &  &  &  &  &  \\ 
   \hline
m & \mpiscatter &  &  &  &  &  &  & \scalebox{0.7}{\textbullet} &  &  &  &  &  &  &  &  &  &  &  &  &  &  &  \\ 
   \hline
s & \mpiallgather &  &  &  &  &  &  &  & \scalebox{0.7}{\textbullet} & \scalebox{0.7}{\textbullet} & \scalebox{0.7}{\textbullet} & \scalebox{0.7}{\textbullet} & \scalebox{0.7}{\textbullet} &  &  & \scalebox{0.7}{\textbullet} & \scalebox{0.7}{\textbullet} &  &  &  &  &  &  \\ 
   \hline
s & \mpiallreduce &  &  &  &  &  &  &  &  &  &  & \scalebox{0.7}{\textbullet} &  &  &  &  &  &  &  &  &  &  &  \\ 
   \hline
s & \mpibcast &  &  &  &  &  &  &  &  &  &  &  &  &  &  &  &  &  &  &  & \scalebox{0.7}{\textbullet} &  & \scalebox{0.7}{\textbullet} \\ 
   \hline
s & \mpireduce &  &  &  &  &  &  &  &  &  &  &  &  &  &  &  &  &  &  & \scalebox{0.7}{\textbullet} & \scalebox{0.7}{\textbullet} & \scalebox{0.7}{\textbullet} &  \\ 
   \hline
s & \mpireducescatterblock &  &  &  &  &  &  &  &  &  &  &  &  &  & \scalebox{0.7}{\textbullet} & \scalebox{0.7}{\textbullet} & \scalebox{0.7}{\textbullet} & \scalebox{0.7}{\textbullet} & \scalebox{0.7}{\textbullet} & \scalebox{0.7}{\textbullet} & \scalebox{0.7}{\textbullet} & \scalebox{0.7}{\textbullet} & \scalebox{0.7}{\textbullet} \\ 
   \hline
s & \mpiscatter &  &  &  &  &  & \scalebox{0.7}{\textbullet} & \scalebox{0.7}{\textbullet} &  &  &  &  &  &  &  &  &  &  &  &  &  &  &  \\ 
   \hline
p & \mpiallgather \guidelt Allreduce &  &  &  &  &  &  &  & \scalebox{0.7}{\textbullet} & \scalebox{0.7}{\textbullet} &  &  &  &  &  &  &  &  &  &  &  &  &  \\ 
   \hline
p & \mpiallgather \guidelt Alltoall &  &  &  &  &  &  &  & \scalebox{0.7}{\textbullet} & \scalebox{0.7}{\textbullet} &  &  &  &  &  &  &  &  &  &  &  &  &  \\ 
   \hline
p & \mpiallreduce \guidelt Reduce$+$Bcast &  &  &  &  &  &  &  &  &  &  & \scalebox{0.7}{\textbullet} & \scalebox{0.7}{\textbullet} & \scalebox{0.7}{\textbullet} &  &  &  &  &  &  &  &  &  \\ 
   \hline
p & \mpiallreduce \guidelt Reduce\_scatter\_block$+$Allgather &  &  &  &  &  &  &  &  &  &  & \scalebox{0.7}{\textbullet} & \scalebox{0.7}{\textbullet} &  & \scalebox{0.7}{\textbullet} & \scalebox{0.7}{\textbullet} &  & \scalebox{0.7}{\textbullet} &  & \scalebox{0.7}{\textbullet} &  & \scalebox{0.7}{\textbullet} &  \\ 
   \hline
p & \mpibcast \guidelt Scatter$+$Allgather &  &  &  &  &  &  &  &  &  &  &  &  & \scalebox{0.7}{\textbullet} & \scalebox{0.7}{\textbullet} & \scalebox{0.7}{\textbullet} & \scalebox{0.7}{\textbullet} &  &  &  &  &  & \scalebox{0.7}{\textbullet} \\ 
   \hline
p & \mpigather \guidelt Allgather & \scalebox{0.7}{\textbullet} & \scalebox{0.7}{\textbullet} & \scalebox{0.7}{\textbullet} & \scalebox{0.7}{\textbullet} & \scalebox{0.7}{\textbullet} & \scalebox{0.7}{\textbullet} & \scalebox{0.7}{\textbullet} &  &  &  &  &  &  &  &  &  &  &  &  &  &  &  \\ 
   \hline
p & \mpigather \guidelt Reduce & \scalebox{0.7}{\textbullet} & \scalebox{0.7}{\textbullet} & \scalebox{0.7}{\textbullet} & \scalebox{0.7}{\textbullet} & \scalebox{0.7}{\textbullet} & \scalebox{0.7}{\textbullet} &  &  &  &  &  &  &  &  &  &  &  &  &  &  &  &  \\ 
   \hline
p & \mpireduce \guidelt Allreduce & \scalebox{0.7}{\textbullet} & \scalebox{0.7}{\textbullet} & \scalebox{0.7}{\textbullet} &  &  &  &  &  &  &  &  &  &  &  & \scalebox{0.7}{\textbullet} & \scalebox{0.7}{\textbullet} &  & \scalebox{0.7}{\textbullet} & \scalebox{0.7}{\textbullet} & \scalebox{0.7}{\textbullet} & \scalebox{0.7}{\textbullet} &  \\ 
   \hline
p & \mpireduce \guidelt Reduce\_scatter\_block$+$Gather &  &  &  &  &  &  &  &  &  &  &  &  &  &  &  &  &  &  & \scalebox{0.7}{\textbullet} & \scalebox{0.7}{\textbullet} & \scalebox{0.7}{\textbullet} &  \\ 
   \hline
p & \mpireducescatter \guidelt Reduce$+$Scatterv &  &  &  &  &  &  &  &  &  &  &  &  &  &  &  &  &  & \scalebox{0.7}{\textbullet} & \scalebox{0.7}{\textbullet} &  &  & \scalebox{0.7}{\textbullet} \\ 
   \hline
p & \mpireducescatterblock \guidelt Reduce+Scatter &  &  &  &  &  &  &  &  &  &  &  &  &  &  & \scalebox{0.7}{\textbullet} & \scalebox{0.7}{\textbullet} & \scalebox{0.7}{\textbullet} & \scalebox{0.7}{\textbullet} & \scalebox{0.7}{\textbullet} & \scalebox{0.7}{\textbullet} & \scalebox{0.7}{\textbullet} & \scalebox{0.7}{\textbullet} \\ 
   \hline
p & \mpiscan \guidelt Exscan$+$Reduce\_local &  &  &  &  &  &  &  &  &  &  &  &  &  &  &  &  & \scalebox{0.7}{\textbullet} & \scalebox{0.7}{\textbullet} &  &  & \scalebox{0.7}{\textbullet} & \scalebox{0.7}{\textbullet} \\ 
   \hline
p & \mpiscatter \guidelt Bcast &  &  &  &  &  & \scalebox{0.7}{\textbullet} & \scalebox{0.7}{\textbullet} &  &  &  &  &  &  &  &  &  &  &  &  &  &  &  \\ 
   \hline
\end{tabular}

  \end{scriptsize}
\end{table}

\begin{table}[!h]
  \centering
  \caption{\label{tab:vsc_case3}Performance-guideline violations of \vscintelmpi using \num{64x16} processes on \machtwo ($\nmpiruns=10$);  violation types: \textbf{m}onotony, \textbf{s}plit-robustness, \textbf{p}attern.}
  \begin{scriptsize}
\begin{tabular}{clacacacacacacacacacaca}
  \hline
type & function & \begin{sideways}\num{1}\end{sideways} & \begin{sideways}\num{2}\end{sideways} & \begin{sideways}\num{4}\end{sideways} & \begin{sideways}\num{8}\end{sideways} & \begin{sideways}\num{16}\end{sideways} & \begin{sideways}\num{32}\end{sideways} & \begin{sideways}\num{64}\end{sideways} & \begin{sideways}\num{100}\end{sideways} & \begin{sideways}\num{128}\end{sideways} & \begin{sideways}\num{256}\end{sideways} & \begin{sideways}\num{512}\end{sideways} & \begin{sideways}\num{1024}\end{sideways} & \begin{sideways}\num{1500}\end{sideways} & \begin{sideways}\num{2048}\end{sideways} & \begin{sideways}\num{4096}\end{sideways} & \begin{sideways}\num{5000}\end{sideways} & \begin{sideways}\num{8192}\end{sideways} & \begin{sideways}\num{10000}\end{sideways} & \begin{sideways}\num{16384}\end{sideways} & \begin{sideways}\num{32768}\end{sideways} & \begin{sideways}\num{102400}\end{sideways} \\ 
  \hline
m & \mpiallreduce &  &  & \scalebox{0.7}{\textbullet} &  & \scalebox{0.7}{\textbullet} &  &  &  &  &  &  & \scalebox{0.7}{\textbullet} &  & \scalebox{0.7}{\textbullet} & \scalebox{0.7}{\textbullet} &  &  &  &  &  &  \\ 
   \hline
m & \mpibcast &  &  &  &  &  &  &  &  &  &  &  &  &  &  &  &  &  & \scalebox{0.7}{\textbullet} &  &  & \scalebox{0.7}{\textbullet} \\ 
   \hline
m & \mpigather &  &  & \scalebox{0.7}{\textbullet} &  &  &  & \scalebox{0.7}{\textbullet} &  & \scalebox{0.7}{\textbullet} &  &  &  &  &  &  &  &  &  &  &  & \scalebox{0.7}{\textbullet} \\ 
   \hline
m & \mpireduce &  &  & \scalebox{0.7}{\textbullet} & \scalebox{0.7}{\textbullet} &  &  &  &  &  &  &  &  &  &  &  &  & \scalebox{0.7}{\textbullet} &  &  &  &  \\ 
   \hline
m & \mpiscan &  &  &  &  &  &  &  & \scalebox{0.7}{\textbullet} &  & \scalebox{0.7}{\textbullet} &  &  &  &  &  &  &  &  &  &  &  \\ 
   \hline
m & \mpiscatter &  &  &  &  &  &  & \scalebox{0.7}{\textbullet} &  &  &  &  &  &  &  &  & \scalebox{0.7}{\textbullet} &  &  &  &  &  \\ 
   \hline
s & \mpiallgather &  &  &  &  &  &  &  &  & \scalebox{0.7}{\textbullet} & \scalebox{0.7}{\textbullet} & \scalebox{0.7}{\textbullet} & \scalebox{0.7}{\textbullet} & \scalebox{0.7}{\textbullet} & \scalebox{0.7}{\textbullet} & \scalebox{0.7}{\textbullet} & \scalebox{0.7}{\textbullet} & \scalebox{0.7}{\textbullet} & \scalebox{0.7}{\textbullet} & \scalebox{0.7}{\textbullet} & \scalebox{0.7}{\textbullet} & \scalebox{0.7}{\textbullet} \\ 
   \hline
s & \mpiallreduce &  &  &  &  &  &  &  &  &  &  & \scalebox{0.7}{\textbullet} & \scalebox{0.7}{\textbullet} &  &  &  &  &  &  &  &  &  \\ 
   \hline
s & \mpibcast &  &  &  &  &  &  &  &  &  &  &  &  &  &  &  &  &  &  &  & \scalebox{0.7}{\textbullet} &  \\ 
   \hline
s & \mpigather &  &  &  &  &  &  &  &  &  &  &  &  &  &  & \scalebox{0.7}{\textbullet} &  & \scalebox{0.7}{\textbullet} & \scalebox{0.7}{\textbullet} & \scalebox{0.7}{\textbullet} & \scalebox{0.7}{\textbullet} &  \\ 
   \hline
s & \mpireduce &  &  &  &  &  &  &  &  &  &  &  &  &  &  &  &  &  &  & \scalebox{0.7}{\textbullet} & \scalebox{0.7}{\textbullet} & \scalebox{0.7}{\textbullet} \\ 
   \hline
s & \mpireducescatterblock &  &  &  &  &  &  &  &  &  &  & \scalebox{0.7}{\textbullet} & \scalebox{0.7}{\textbullet} & \scalebox{0.7}{\textbullet} & \scalebox{0.7}{\textbullet} & \scalebox{0.7}{\textbullet} & \scalebox{0.7}{\textbullet} & \scalebox{0.7}{\textbullet} & \scalebox{0.7}{\textbullet} & \scalebox{0.7}{\textbullet} & \scalebox{0.7}{\textbullet} & \scalebox{0.7}{\textbullet} \\ 
   \hline
s & \mpiscatter &  &  &  &  &  & \scalebox{0.7}{\textbullet} & \scalebox{0.7}{\textbullet} & \scalebox{0.7}{\textbullet} & \scalebox{0.7}{\textbullet} &  &  &  &  &  &  &  &  &  &  &  &  \\ 
   \hline
p & \mpiallreduce \guidelt Reduce$+$Bcast &  &  &  &  &  &  &  &  &  &  & \scalebox{0.7}{\textbullet} & \scalebox{0.7}{\textbullet} & \scalebox{0.7}{\textbullet} &  &  &  &  &  &  &  &  \\ 
   \hline
p & \mpiallreduce \guidelt Reduce\_scatter\_block$+$Allgather &  &  &  &  &  &  &  &  &  &  &  & \scalebox{0.7}{\textbullet} &  & \scalebox{0.7}{\textbullet} & \scalebox{0.7}{\textbullet} &  & \scalebox{0.7}{\textbullet} &  & \scalebox{0.7}{\textbullet} & \scalebox{0.7}{\textbullet} & \scalebox{0.7}{\textbullet} \\ 
   \hline
p & \mpibcast \guidelt Scatter$+$Allgather &  &  &  &  &  &  &  &  &  &  &  &  & \scalebox{0.7}{\textbullet} & \scalebox{0.7}{\textbullet} & \scalebox{0.7}{\textbullet} & \scalebox{0.7}{\textbullet} & \scalebox{0.7}{\textbullet} &  &  & \scalebox{0.7}{\textbullet} &  \\ 
   \hline
p & \mpigather \guidelt Allgather & \scalebox{0.7}{\textbullet} & \scalebox{0.7}{\textbullet} & \scalebox{0.7}{\textbullet} & \scalebox{0.7}{\textbullet} & \scalebox{0.7}{\textbullet} & \scalebox{0.7}{\textbullet} & \scalebox{0.7}{\textbullet} & \scalebox{0.7}{\textbullet} & \scalebox{0.7}{\textbullet} & \scalebox{0.7}{\textbullet} &  &  &  &  &  &  &  &  &  &  &  \\ 
   \hline
p & \mpigather \guidelt Reduce & \scalebox{0.7}{\textbullet} & \scalebox{0.7}{\textbullet} & \scalebox{0.7}{\textbullet} & \scalebox{0.7}{\textbullet} & \scalebox{0.7}{\textbullet} &  &  &  &  &  &  &  &  &  &  &  &  &  &  &  &  \\ 
   \hline
p & \mpireduce \guidelt Allreduce & \scalebox{0.7}{\textbullet} & \scalebox{0.7}{\textbullet} & \scalebox{0.7}{\textbullet} &  &  &  &  &  &  &  &  &  &  &  & \scalebox{0.7}{\textbullet} & \scalebox{0.7}{\textbullet} &  & \scalebox{0.7}{\textbullet} & \scalebox{0.7}{\textbullet} & \scalebox{0.7}{\textbullet} & \scalebox{0.7}{\textbullet} \\ 
   \hline
p & \mpireduce \guidelt Reduce\_scatter\_block$+$Gather &  &  &  &  &  &  &  &  &  &  &  &  &  &  &  &  &  &  &  & \scalebox{0.7}{\textbullet} & \scalebox{0.7}{\textbullet} \\ 
   \hline
p & \mpireducescatter \guidelt Reduce$+$Scatterv &  &  &  &  &  &  &  &  &  &  &  &  &  &  &  & \scalebox{0.7}{\textbullet} &  & \scalebox{0.7}{\textbullet} &  & \scalebox{0.7}{\textbullet} & \scalebox{0.7}{\textbullet} \\ 
   \hline
p & \mpireducescatterblock \guidelt Reduce+Scatter &  &  &  &  &  &  &  &  &  &  &  &  &  &  & \scalebox{0.7}{\textbullet} & \scalebox{0.7}{\textbullet} & \scalebox{0.7}{\textbullet} & \scalebox{0.7}{\textbullet} & \scalebox{0.7}{\textbullet} & \scalebox{0.7}{\textbullet} & \scalebox{0.7}{\textbullet} \\ 
   \hline
p & \mpiscan \guidelt Exscan$+$Reduce\_local &  &  &  &  &  &  & \scalebox{0.7}{\textbullet} &  & \scalebox{0.7}{\textbullet} &  &  &  &  &  &  &  & \scalebox{0.7}{\textbullet} & \scalebox{0.7}{\textbullet} & \scalebox{0.7}{\textbullet} & \scalebox{0.7}{\textbullet} &  \\ 
   \hline
p & \mpiscatter \guidelt Bcast &  &  &  &  &  &  &  & \scalebox{0.7}{\textbullet} &  &  &  &  &  &  &  &  &  &  &  &  &  \\ 
   \hline
\end{tabular}

  \end{scriptsize}
\end{table}

\FloatBarrier

\clearpage

\subsection{\machthree}

\begin{table}[!h]
  \centering
  \caption{\label{tab:fermi_gl_check_overview}Performance-guideline violations of \fermibgqmpi for different process configurations on \machthree; violation types: monotony, split-robustness, pattern; message sizes between \SI{1}{\byte} and \SI{100}{\kibi\byte}.}
  \begin{scriptsize}
\begin{tabular}{ccr}
  \toprule
\#processes & type & {\tiny{}Fermi} \\ 
  \midrule
64x16 & m & 4/9 \\ 
   \midrule
64x16 & s & 3/9 \\ 
   \midrule
64x16 & p & 12/15 \\ 
   \midrule
256x16 & m & 4/9 \\ 
   \midrule
256x16 & s & 4/9 \\ 
   \midrule
256x16 & p & 11/15 \\ 
   \midrule
1024x16 & m & 5/9 \\ 
   \midrule
1024x16 & s & 4/9 \\ 
   \midrule
1024x16 & p & 9/15 \\ 
   \bottomrule
\end{tabular}

  \end{scriptsize}
\end{table}

\begin{table}[!h]
  \centering
  \caption{\label{tab:fermi_gl_patterns}Pattern guideline violations of \fermibgqmpi for different process configurations on \machthree; message sizes between \SI{1}{\byte} and \SI{100}{\kibi\byte}.}
  \begin{scriptsize}
\begin{tabular}{lccc}
  \toprule
guideline & {\tiny{}64x16} & {\tiny{}256x16} & {\tiny{}1024x16} \\ 
  \midrule
\mpiallgather \guidelt Allreduce &  &  &  \\ 
   \midrule
\mpiallgather \guidelt Alltoall & \scalebox{0.7}{\textbullet} & \scalebox{0.7}{\textbullet} &  \\ 
   \midrule
\mpiallgather \guidelt Gather$+$Bcast & \scalebox{0.7}{\textbullet} & \scalebox{0.7}{\textbullet} & \scalebox{0.7}{\textbullet} \\ 
   \midrule
\mpiallreduce \guidelt Reduce$+$Bcast & \scalebox{0.7}{\textbullet} & \scalebox{0.7}{\textbullet} & \scalebox{0.7}{\textbullet} \\ 
   \midrule
\mpiallreduce \guidelt Reduce\_scatter\_block$+$Allgather & \scalebox{0.7}{\textbullet} & \scalebox{0.7}{\textbullet} & \scalebox{0.7}{\textbullet} \\ 
   \midrule
\mpibcast \guidelt Scatter$+$Allgather & \scalebox{0.7}{\textbullet} &  &  \\ 
   \midrule
\mpigather \guidelt Allgather & \scalebox{0.7}{\textbullet} & \scalebox{0.7}{\textbullet} &  \\ 
   \midrule
\mpigather \guidelt Reduce &  &  &  \\ 
   \midrule
\mpireduce \guidelt Allreduce & \scalebox{0.7}{\textbullet} & \scalebox{0.7}{\textbullet} & \scalebox{0.7}{\textbullet} \\ 
   \midrule
\mpireduce \guidelt Reduce\_scatter\_block$+$Gather & \scalebox{0.7}{\textbullet} & \scalebox{0.7}{\textbullet} & \scalebox{0.7}{\textbullet} \\ 
   \midrule
\mpireducescatterblock \guidelt Reduce+Scatter & \scalebox{0.7}{\textbullet} & \scalebox{0.7}{\textbullet} & \scalebox{0.7}{\textbullet} \\ 
   \midrule
\mpireducescatter \guidelt Allreduce & \scalebox{0.7}{\textbullet} & \scalebox{0.7}{\textbullet} & \scalebox{0.7}{\textbullet} \\ 
   \midrule
\mpireducescatter \guidelt Reduce$+$Scatterv & \scalebox{0.7}{\textbullet} & \scalebox{0.7}{\textbullet} & \scalebox{0.7}{\textbullet} \\ 
   \midrule
\mpiscan \guidelt Exscan$+$Reduce\_local &  &  &  \\ 
   \midrule
\mpiscatter \guidelt Bcast & \scalebox{0.7}{\textbullet} & \scalebox{0.7}{\textbullet} & \scalebox{0.7}{\textbullet} \\ 
   \bottomrule
\end{tabular}

  \end{scriptsize}
\end{table}

\begin{table}[!h]
  \centering
  \caption{\label{tab:fermi_case1}Performance-guideline violations of \fermibgqmpi using \num{64x16} processes on \machthree ($\nmpiruns=10$); violation types: \textbf{m}onotony, \textbf{s}plit-robustness, \textbf{p}attern.}
  \begin{scriptsize}
\begin{tabular}{clacacacacacacacacacaca}
  \hline
type & function & \begin{sideways}\num{1}\end{sideways} & \begin{sideways}\num{2}\end{sideways} & \begin{sideways}\num{4}\end{sideways} & \begin{sideways}\num{8}\end{sideways} & \begin{sideways}\num{16}\end{sideways} & \begin{sideways}\num{32}\end{sideways} & \begin{sideways}\num{64}\end{sideways} & \begin{sideways}\num{100}\end{sideways} & \begin{sideways}\num{128}\end{sideways} & \begin{sideways}\num{256}\end{sideways} & \begin{sideways}\num{512}\end{sideways} & \begin{sideways}\num{1024}\end{sideways} & \begin{sideways}\num{1500}\end{sideways} & \begin{sideways}\num{2048}\end{sideways} & \begin{sideways}\num{4096}\end{sideways} & \begin{sideways}\num{5000}\end{sideways} & \begin{sideways}\num{8192}\end{sideways} & \begin{sideways}\num{10000}\end{sideways} & \begin{sideways}\num{16384}\end{sideways} & \begin{sideways}\num{32768}\end{sideways} & \begin{sideways}\num{102400}\end{sideways} \\ 
  \hline
m & \mpiallgather &  &  & \scalebox{0.7}{\textbullet} &  &  &  &  &  &  &  &  &  &  & \scalebox{0.7}{\textbullet} &  &  &  &  &  &  &  \\ 
   \hline
m & \mpiallreduce &  &  &  &  &  &  &  &  &  &  &  & \scalebox{0.7}{\textbullet} &  &  & \scalebox{0.7}{\textbullet} &  & \scalebox{0.7}{\textbullet} &  & \scalebox{0.7}{\textbullet} &  &  \\ 
   \hline
m & \mpigather &  &  & \scalebox{0.7}{\textbullet} &  &  &  & \scalebox{0.7}{\textbullet} &  &  &  &  &  &  & \scalebox{0.7}{\textbullet} &  &  & \scalebox{0.7}{\textbullet} &  & \scalebox{0.7}{\textbullet} &  &  \\ 
   \hline
m & \mpireduce &  &  &  &  &  &  &  &  &  &  &  & \scalebox{0.7}{\textbullet} & \scalebox{0.7}{\textbullet} &  & \scalebox{0.7}{\textbullet} &  & \scalebox{0.7}{\textbullet} &  & \scalebox{0.7}{\textbullet} &  &  \\ 
   \hline
s & \mpiallgather &  &  &  &  &  &  &  &  &  &  &  &  & \scalebox{0.7}{\textbullet} &  &  &  &  &  &  &  &  \\ 
   \hline
s & \mpireducescatter &  &  &  &  &  &  &  &  &  &  & \scalebox{0.7}{\textbullet} &  &  &  &  &  &  &  &  &  &  \\ 
   \hline
s & \mpireducescatterblock &  &  &  &  &  &  &  &  &  &  & \scalebox{0.7}{\textbullet} &  &  &  &  &  &  &  &  &  &  \\ 
   \hline
p & \mpiallgather \guidelt Alltoall &  &  &  &  &  &  &  &  &  &  &  &  & \scalebox{0.7}{\textbullet} &  &  &  &  &  &  &  &  \\ 
   \hline
p & \mpiallgather \guidelt Gather$+$Bcast & \scalebox{0.7}{\textbullet} & \scalebox{0.7}{\textbullet} &  &  &  &  & \scalebox{0.7}{\textbullet} & \scalebox{0.7}{\textbullet} & \scalebox{0.7}{\textbullet} & \scalebox{0.7}{\textbullet} & \scalebox{0.7}{\textbullet} & \scalebox{0.7}{\textbullet} & \scalebox{0.7}{\textbullet} & \scalebox{0.7}{\textbullet} & \scalebox{0.7}{\textbullet} &  & \scalebox{0.7}{\textbullet} &  &  &  &  \\ 
   \hline
p & \mpiallreduce \guidelt Reduce$+$Bcast &  &  &  &  &  &  &  &  &  &  &  &  & \scalebox{0.7}{\textbullet} & \scalebox{0.7}{\textbullet} & \scalebox{0.7}{\textbullet} & \scalebox{0.7}{\textbullet} & \scalebox{0.7}{\textbullet} & \scalebox{0.7}{\textbullet} & \scalebox{0.7}{\textbullet} & \scalebox{0.7}{\textbullet} &  \\ 
   \hline
p & \mpiallreduce \guidelt Reduce\_scatter\_block$+$Allgather &  &  &  &  &  &  &  &  &  &  & \scalebox{0.7}{\textbullet} &  &  & \scalebox{0.7}{\textbullet} & \scalebox{0.7}{\textbullet} &  & \scalebox{0.7}{\textbullet} &  &  &  &  \\ 
   \hline
p & \mpibcast \guidelt Scatter$+$Allgather &  &  &  &  &  &  &  &  &  &  & \scalebox{0.7}{\textbullet} &  &  &  &  &  &  &  &  &  &  \\ 
   \hline
p & \mpigather \guidelt Allgather &  &  & \scalebox{0.7}{\textbullet} & \scalebox{0.7}{\textbullet} & \scalebox{0.7}{\textbullet} & \scalebox{0.7}{\textbullet} &  &  &  &  &  &  &  &  &  &  &  &  &  &  &  \\ 
   \hline
p & \mpireduce \guidelt Allreduce & \scalebox{0.7}{\textbullet} & \scalebox{0.7}{\textbullet} & \scalebox{0.7}{\textbullet} & \scalebox{0.7}{\textbullet} & \scalebox{0.7}{\textbullet} & \scalebox{0.7}{\textbullet} & \scalebox{0.7}{\textbullet} & \scalebox{0.7}{\textbullet} & \scalebox{0.7}{\textbullet} & \scalebox{0.7}{\textbullet} & \scalebox{0.7}{\textbullet} & \scalebox{0.7}{\textbullet} &  &  &  &  &  &  &  &  &  \\ 
   \hline
p & \mpireduce \guidelt Reduce\_scatter\_block$+$Gather &  &  &  &  &  &  &  &  &  &  & \scalebox{0.7}{\textbullet} & \scalebox{0.7}{\textbullet} &  & \scalebox{0.7}{\textbullet} &  &  &  &  &  &  &  \\ 
   \hline
p & \mpireducescatterblock \guidelt Reduce+Scatter &  &  &  &  &  & \scalebox{0.7}{\textbullet} & \scalebox{0.7}{\textbullet} & \scalebox{0.7}{\textbullet} & \scalebox{0.7}{\textbullet} & \scalebox{0.7}{\textbullet} & \scalebox{0.7}{\textbullet} & \scalebox{0.7}{\textbullet} & \scalebox{0.7}{\textbullet} & \scalebox{0.7}{\textbullet} & \scalebox{0.7}{\textbullet} & \scalebox{0.7}{\textbullet} & \scalebox{0.7}{\textbullet} & \scalebox{0.7}{\textbullet} & \scalebox{0.7}{\textbullet} & \scalebox{0.7}{\textbullet} & \scalebox{0.7}{\textbullet} \\ 
   \hline
p & \mpireducescatter \guidelt Allreduce &  &  &  &  &  & \scalebox{0.7}{\textbullet} & \scalebox{0.7}{\textbullet} & \scalebox{0.7}{\textbullet} & \scalebox{0.7}{\textbullet} & \scalebox{0.7}{\textbullet} & \scalebox{0.7}{\textbullet} & \scalebox{0.7}{\textbullet} & \scalebox{0.7}{\textbullet} & \scalebox{0.7}{\textbullet} & \scalebox{0.7}{\textbullet} & \scalebox{0.7}{\textbullet} & \scalebox{0.7}{\textbullet} & \scalebox{0.7}{\textbullet} & \scalebox{0.7}{\textbullet} & \scalebox{0.7}{\textbullet} & \scalebox{0.7}{\textbullet} \\ 
   \hline
p & \mpireducescatter \guidelt Reduce$+$Scatterv &  &  &  &  &  &  &  &  &  &  & \scalebox{0.7}{\textbullet} & \scalebox{0.7}{\textbullet} & \scalebox{0.7}{\textbullet} & \scalebox{0.7}{\textbullet} & \scalebox{0.7}{\textbullet} & \scalebox{0.7}{\textbullet} & \scalebox{0.7}{\textbullet} & \scalebox{0.7}{\textbullet} & \scalebox{0.7}{\textbullet} & \scalebox{0.7}{\textbullet} & \scalebox{0.7}{\textbullet} \\ 
   \hline
p & \mpiscatter \guidelt Bcast & \scalebox{0.7}{\textbullet} & \scalebox{0.7}{\textbullet} & \scalebox{0.7}{\textbullet} & \scalebox{0.7}{\textbullet} & \scalebox{0.7}{\textbullet} & \scalebox{0.7}{\textbullet} & \scalebox{0.7}{\textbullet} & \scalebox{0.7}{\textbullet} & \scalebox{0.7}{\textbullet} & \scalebox{0.7}{\textbullet} & \scalebox{0.7}{\textbullet} & \scalebox{0.7}{\textbullet} & \scalebox{0.7}{\textbullet} & \scalebox{0.7}{\textbullet} &  &  &  &  &  &  &  \\ 
   \hline
\end{tabular}

  \end{scriptsize}
\end{table}

\begin{table}[!h]
  \centering
  \caption{\label{tab:fermi_case3}Performance-guideline violations of \fermibgqmpi using \num{256x16} processes on \machthree ($\nmpiruns=5$); violation types: \textbf{m}onotony, \textbf{s}plit-robustness, \textbf{p}attern.}
  \begin{scriptsize}
\begin{tabular}{clacacacacacacacac}
  \hline
type & function & \begin{sideways}\num{1}\end{sideways} & \begin{sideways}\num{2}\end{sideways} & \begin{sideways}\num{4}\end{sideways} & \begin{sideways}\num{8}\end{sideways} & \begin{sideways}\num{16}\end{sideways} & \begin{sideways}\num{32}\end{sideways} & \begin{sideways}\num{64}\end{sideways} & \begin{sideways}\num{100}\end{sideways} & \begin{sideways}\num{128}\end{sideways} & \begin{sideways}\num{256}\end{sideways} & \begin{sideways}\num{512}\end{sideways} & \begin{sideways}\num{1024}\end{sideways} & \begin{sideways}\num{1500}\end{sideways} & \begin{sideways}\num{2048}\end{sideways} & \begin{sideways}\num{4096}\end{sideways} & \begin{sideways}\num{8192}\end{sideways} \\ 
  \hline
m & \mpiallgather &  &  & \scalebox{0.7}{\textbullet} &  &  &  &  &  &  &  &  &  &  & \scalebox{0.7}{\textbullet} &  &  \\ 
   \hline
m & \mpiallreduce &  &  &  &  &  &  &  &  &  &  &  & \scalebox{0.7}{\textbullet} &  &  & \scalebox{0.7}{\textbullet} &  \\ 
   \hline
m & \mpigather &  &  & \scalebox{0.7}{\textbullet} &  &  &  &  &  &  &  &  &  &  & \scalebox{0.7}{\textbullet} &  &  \\ 
   \hline
m & \mpireduce &  &  &  &  &  &  &  &  &  &  &  & \scalebox{0.7}{\textbullet} & \scalebox{0.7}{\textbullet} &  &  &  \\ 
   \hline
s & \mpiallgather &  &  &  &  &  &  &  &  &  &  & \scalebox{0.7}{\textbullet} &  &  &  &  &  \\ 
   \hline
s & \mpigather &  &  &  &  &  &  &  &  &  &  &  &  & \scalebox{0.7}{\textbullet} &  &  &  \\ 
   \hline
s & \mpireducescatter &  &  &  &  &  &  &  &  & \scalebox{0.7}{\textbullet} & \scalebox{0.7}{\textbullet} & \scalebox{0.7}{\textbullet} &  &  &  &  &  \\ 
   \hline
s & \mpireducescatterblock &  &  &  &  &  &  &  &  & \scalebox{0.7}{\textbullet} & \scalebox{0.7}{\textbullet} & \scalebox{0.7}{\textbullet} &  &  &  &  &  \\ 
   \hline
p & \mpiallgather \guidelt Alltoall &  &  &  &  &  &  &  &  &  &  & \scalebox{0.7}{\textbullet} &  & \scalebox{0.7}{\textbullet} &  &  &  \\ 
   \hline
p & \mpiallgather \guidelt Gather$+$Bcast & \scalebox{0.7}{\textbullet} & \scalebox{0.7}{\textbullet} &  &  & \scalebox{0.7}{\textbullet} & \scalebox{0.7}{\textbullet} & \scalebox{0.7}{\textbullet} & \scalebox{0.7}{\textbullet} & \scalebox{0.7}{\textbullet} & \scalebox{0.7}{\textbullet} & \scalebox{0.7}{\textbullet} & \scalebox{0.7}{\textbullet} & \scalebox{0.7}{\textbullet} & \scalebox{0.7}{\textbullet} & \scalebox{0.7}{\textbullet} & \scalebox{0.7}{\textbullet} \\ 
   \hline
p & \mpiallreduce \guidelt Reduce$+$Bcast &  &  &  &  &  &  &  &  &  &  &  &  & \scalebox{0.7}{\textbullet} & \scalebox{0.7}{\textbullet} & \scalebox{0.7}{\textbullet} & \scalebox{0.7}{\textbullet} \\ 
   \hline
p & \mpiallreduce \guidelt Reduce\_scatter\_block$+$Allgather &  &  &  &  &  &  &  &  &  &  & \scalebox{0.7}{\textbullet} & \scalebox{0.7}{\textbullet} &  & \scalebox{0.7}{\textbullet} &  &  \\ 
   \hline
p & \mpigather \guidelt Allgather &  &  & \scalebox{0.7}{\textbullet} & \scalebox{0.7}{\textbullet} &  &  &  &  &  &  &  &  &  &  &  &  \\ 
   \hline
p & \mpireduce \guidelt Allreduce & \scalebox{0.7}{\textbullet} & \scalebox{0.7}{\textbullet} & \scalebox{0.7}{\textbullet} & \scalebox{0.7}{\textbullet} &  & \scalebox{0.7}{\textbullet} & \scalebox{0.7}{\textbullet} & \scalebox{0.7}{\textbullet} & \scalebox{0.7}{\textbullet} & \scalebox{0.7}{\textbullet} & \scalebox{0.7}{\textbullet} & \scalebox{0.7}{\textbullet} &  &  &  &  \\ 
   \hline
p & \mpireduce \guidelt Reduce\_scatter\_block$+$Gather &  &  &  &  &  &  &  &  &  &  & \scalebox{0.7}{\textbullet} & \scalebox{0.7}{\textbullet} &  & \scalebox{0.7}{\textbullet} &  &  \\ 
   \hline
p & \mpireducescatterblock \guidelt Reduce+Scatter &  &  &  & \scalebox{0.7}{\textbullet} & \scalebox{0.7}{\textbullet} & \scalebox{0.7}{\textbullet} & \scalebox{0.7}{\textbullet} & \scalebox{0.7}{\textbullet} & \scalebox{0.7}{\textbullet} & \scalebox{0.7}{\textbullet} & \scalebox{0.7}{\textbullet} & \scalebox{0.7}{\textbullet} & \scalebox{0.7}{\textbullet} & \scalebox{0.7}{\textbullet} & \scalebox{0.7}{\textbullet} & \scalebox{0.7}{\textbullet} \\ 
   \hline
p & \mpireducescatter \guidelt Allreduce &  &  & \scalebox{0.7}{\textbullet} & \scalebox{0.7}{\textbullet} & \scalebox{0.7}{\textbullet} & \scalebox{0.7}{\textbullet} & \scalebox{0.7}{\textbullet} & \scalebox{0.7}{\textbullet} & \scalebox{0.7}{\textbullet} & \scalebox{0.7}{\textbullet} & \scalebox{0.7}{\textbullet} & \scalebox{0.7}{\textbullet} & \scalebox{0.7}{\textbullet} & \scalebox{0.7}{\textbullet} & \scalebox{0.7}{\textbullet} & \scalebox{0.7}{\textbullet} \\ 
   \hline
p & \mpireducescatter \guidelt Reduce$+$Scatterv &  &  &  &  &  &  &  &  & \scalebox{0.7}{\textbullet} & \scalebox{0.7}{\textbullet} & \scalebox{0.7}{\textbullet} & \scalebox{0.7}{\textbullet} & \scalebox{0.7}{\textbullet} & \scalebox{0.7}{\textbullet} & \scalebox{0.7}{\textbullet} & \scalebox{0.7}{\textbullet} \\ 
   \hline
p & \mpiscatter \guidelt Bcast & \scalebox{0.7}{\textbullet} & \scalebox{0.7}{\textbullet} & \scalebox{0.7}{\textbullet} & \scalebox{0.7}{\textbullet} & \scalebox{0.7}{\textbullet} & \scalebox{0.7}{\textbullet} & \scalebox{0.7}{\textbullet} & \scalebox{0.7}{\textbullet} & \scalebox{0.7}{\textbullet} & \scalebox{0.7}{\textbullet} & \scalebox{0.7}{\textbullet} &  &  &  &  &  \\ 
   \hline
\end{tabular}

  \end{scriptsize}
\end{table}

\begin{table}[!h]
  \centering
  \caption{\label{tab:fermi_case4}Performance-guideline violations of \fermibgqmpi using \num{1024x16} processes on \machthree ($\nmpiruns=5$); violation types: \textbf{m}onotony, \textbf{s}plit-robustness, \textbf{p}attern.}
  \begin{scriptsize}
\begin{tabular}{clacacacacacacacac}
  \hline
type & function & \begin{sideways}\num{1}\end{sideways} & \begin{sideways}\num{2}\end{sideways} & \begin{sideways}\num{4}\end{sideways} & \begin{sideways}\num{8}\end{sideways} & \begin{sideways}\num{16}\end{sideways} & \begin{sideways}\num{32}\end{sideways} & \begin{sideways}\num{64}\end{sideways} & \begin{sideways}\num{100}\end{sideways} & \begin{sideways}\num{128}\end{sideways} & \begin{sideways}\num{256}\end{sideways} & \begin{sideways}\num{512}\end{sideways} & \begin{sideways}\num{1024}\end{sideways} & \begin{sideways}\num{1500}\end{sideways} & \begin{sideways}\num{2048}\end{sideways} & \begin{sideways}\num{4096}\end{sideways} & \begin{sideways}\num{8192}\end{sideways} \\ 
  \hline
m & \mpiallgather &  &  & \scalebox{0.7}{\textbullet} &  &  &  &  &  &  & \scalebox{0.7}{\textbullet} &  &  &  & \scalebox{0.7}{\textbullet} &  &  \\ 
   \hline
m & \mpigather &  &  &  &  &  &  &  &  &  &  &  &  &  & \scalebox{0.7}{\textbullet} &  &  \\ 
   \hline
m & \mpireduce &  &  &  &  &  &  &  &  &  &  &  &  & \scalebox{0.7}{\textbullet} &  &  &  \\ 
   \hline
m & \mpireducescatter &  &  &  &  &  &  & \scalebox{0.7}{\textbullet} & \scalebox{0.7}{\textbullet} &  &  &  &  &  &  &  &  \\ 
   \hline
m & \mpireducescatterblock &  &  &  &  &  &  &  & \scalebox{0.7}{\textbullet} &  &  &  &  &  &  &  &  \\ 
   \hline
s & \mpiallgather &  &  &  &  &  &  &  & \scalebox{0.7}{\textbullet} & \scalebox{0.7}{\textbullet} &  &  &  &  &  &  &  \\ 
   \hline
s & \mpigather &  &  &  &  &  &  &  &  &  &  &  &  & \scalebox{0.7}{\textbullet} &  &  &  \\ 
   \hline
s & \mpireducescatter &  &  &  &  &  & \scalebox{0.7}{\textbullet} & \scalebox{0.7}{\textbullet} & \scalebox{0.7}{\textbullet} & \scalebox{0.7}{\textbullet} & \scalebox{0.7}{\textbullet} & \scalebox{0.7}{\textbullet} &  &  &  &  &  \\ 
   \hline
s & \mpireducescatterblock &  &  &  &  &  & \scalebox{0.7}{\textbullet} & \scalebox{0.7}{\textbullet} & \scalebox{0.7}{\textbullet} & \scalebox{0.7}{\textbullet} & \scalebox{0.7}{\textbullet} & \scalebox{0.7}{\textbullet} &  &  &  &  &  \\ 
   \hline
p & \mpiallgather \guidelt Gather$+$Bcast & \scalebox{0.7}{\textbullet} & \scalebox{0.7}{\textbullet} & \scalebox{0.7}{\textbullet} & \scalebox{0.7}{\textbullet} & \scalebox{0.7}{\textbullet} & \scalebox{0.7}{\textbullet} & \scalebox{0.7}{\textbullet} & \scalebox{0.7}{\textbullet} & \scalebox{0.7}{\textbullet} & \scalebox{0.7}{\textbullet} & \scalebox{0.7}{\textbullet} & \scalebox{0.7}{\textbullet} & \scalebox{0.7}{\textbullet} & \scalebox{0.7}{\textbullet} & \scalebox{0.7}{\textbullet} & \scalebox{0.7}{\textbullet} \\ 
   \hline
p & \mpiallreduce \guidelt Reduce$+$Bcast &  &  &  &  &  &  &  &  &  &  &  &  & \scalebox{0.7}{\textbullet} & \scalebox{0.7}{\textbullet} & \scalebox{0.7}{\textbullet} & \scalebox{0.7}{\textbullet} \\ 
   \hline
p & \mpiallreduce \guidelt Reduce\_scatter\_block$+$Allgather &  &  &  &  &  &  &  &  &  &  & \scalebox{0.7}{\textbullet} & \scalebox{0.7}{\textbullet} &  & \scalebox{0.7}{\textbullet} & \scalebox{0.7}{\textbullet} & \scalebox{0.7}{\textbullet} \\ 
   \hline
p & \mpireduce \guidelt Allreduce & \scalebox{0.7}{\textbullet} & \scalebox{0.7}{\textbullet} & \scalebox{0.7}{\textbullet} &  &  &  & \scalebox{0.7}{\textbullet} & \scalebox{0.7}{\textbullet} & \scalebox{0.7}{\textbullet} & \scalebox{0.7}{\textbullet} & \scalebox{0.7}{\textbullet} & \scalebox{0.7}{\textbullet} &  &  &  &  \\ 
   \hline
p & \mpireduce \guidelt Reduce\_scatter\_block$+$Gather &  &  &  &  &  &  &  &  &  &  & \scalebox{0.7}{\textbullet} & \scalebox{0.7}{\textbullet} &  & \scalebox{0.7}{\textbullet} & \scalebox{0.7}{\textbullet} & \scalebox{0.7}{\textbullet} \\ 
   \hline
p & \mpireducescatterblock \guidelt Reduce+Scatter & \scalebox{0.7}{\textbullet} & \scalebox{0.7}{\textbullet} & \scalebox{0.7}{\textbullet} & \scalebox{0.7}{\textbullet} & \scalebox{0.7}{\textbullet} & \scalebox{0.7}{\textbullet} & \scalebox{0.7}{\textbullet} & \scalebox{0.7}{\textbullet} & \scalebox{0.7}{\textbullet} & \scalebox{0.7}{\textbullet} & \scalebox{0.7}{\textbullet} & \scalebox{0.7}{\textbullet} & \scalebox{0.7}{\textbullet} & \scalebox{0.7}{\textbullet} & \scalebox{0.7}{\textbullet} & \scalebox{0.7}{\textbullet} \\ 
   \hline
p & \mpireducescatter \guidelt Allreduce & \scalebox{0.7}{\textbullet} & \scalebox{0.7}{\textbullet} & \scalebox{0.7}{\textbullet} & \scalebox{0.7}{\textbullet} & \scalebox{0.7}{\textbullet} & \scalebox{0.7}{\textbullet} & \scalebox{0.7}{\textbullet} & \scalebox{0.7}{\textbullet} & \scalebox{0.7}{\textbullet} & \scalebox{0.7}{\textbullet} & \scalebox{0.7}{\textbullet} & \scalebox{0.7}{\textbullet} & \scalebox{0.7}{\textbullet} & \scalebox{0.7}{\textbullet} & \scalebox{0.7}{\textbullet} & \scalebox{0.7}{\textbullet} \\ 
   \hline
p & \mpireducescatter \guidelt Reduce$+$Scatterv &  &  &  &  &  & \scalebox{0.7}{\textbullet} & \scalebox{0.7}{\textbullet} & \scalebox{0.7}{\textbullet} & \scalebox{0.7}{\textbullet} & \scalebox{0.7}{\textbullet} & \scalebox{0.7}{\textbullet} & \scalebox{0.7}{\textbullet} & \scalebox{0.7}{\textbullet} & \scalebox{0.7}{\textbullet} & \scalebox{0.7}{\textbullet} & \scalebox{0.7}{\textbullet} \\ 
   \hline
p & \mpiscatter \guidelt Bcast & \scalebox{0.7}{\textbullet} & \scalebox{0.7}{\textbullet} & \scalebox{0.7}{\textbullet} & \scalebox{0.7}{\textbullet} & \scalebox{0.7}{\textbullet} & \scalebox{0.7}{\textbullet} & \scalebox{0.7}{\textbullet} & \scalebox{0.7}{\textbullet} & \scalebox{0.7}{\textbullet} &  &  &  &  &  &  &  \\ 
   \hline
\end{tabular}

  \end{scriptsize}
\end{table}

\end{document}